\documentclass[usenatbib,usegraphicx]{mn2e}
\usepackage{times}
\usepackage{graphicx, epsfig}
\usepackage[fleqn]{amsmath}
\usepackage{amsfonts}
\usepackage{amssymb}
\usepackage{deluxetable}
\usepackage[total={17.8cm,24.0cm},centering]{geometry}
\usepackage{times}

\title[Morphologies, features and alignment of early-type galaxies]{The ATLAS$^{\rm 3D}$ project -- II. Morphologies, kinemetric features and alignment between  photometric and kinematic axes of early-type galaxies}

\author[Davor Krajnovi\'c et al.]  {Davor
   Krajnovi\'c,$^{1}$\thanks{E-mail: dkrajnov@eso.org}, Eric Emsellem$^{1,2}$, Michele Cappellari${^3}$,   Katherine Alatalo$^{4}$, \newauthor Leo Blitz$^{4}$, Maxime Bois$^{1,2}$, Fr\'ed\'eric Bournaud$^{5}$,  Martin Bureau$^{3}$, Roger L. Davies$^{3}$, \newauthor Timothy A. Davis$^{3}$, P. T. de Zeeuw$^{1,6}$, Sadegh Khochfar$^{7}$, Harald Kuntschner$^{8}$, \newauthor Pierre-Yves Lablanche$^{1,2}$, Richard M. McDermid$^{9}$, Raffaella Morganti$^{10,11}$, \newauthor Thorsten Naab$^{12,13}$, Tom Oosterloo$^{10,11}$, Marc Sarzi$^{14}$, Nicholas Scott$^{3}$, Paolo Serra$^{10}$,  \newauthor Anne-Marie Weijmans$^{15}$\thanks{Dunlop Fellow}, and Lisa M. Young$^{16}$  \\
   $^1$European Southern Observatory,Karl-Schwarzschild-Strasse 2, 85748 Garching bei M\"unchen, Germany\\
   $^{2}$  Universit\'e Lyon 1, CRAL, Observatoire de Lyon, 9 avenue Charles Andr\'e, F-69230 Saint-Genis Laval, France \\
   $^{3}$ Sub-department of Astrophysics, University of Oxford, Denys Wilkinson Building, Keble Road, Oxford OX1 3RH\\
   $^{4}$  Department of Astronomy and Radio Astronomy Laboratory, University of California, Berkeley, CA 94720, USA  \\
   $^{5}$  CEA, IRFU, SAp et Laboratoire AIM, CEA Saclay -- CNRS -- Universit\'e Paris Diderot, 91191 Gif-sur-Yvette, France  \\
   $^{6}$  Sterrewacht Leiden, Leiden University, Postbus 9513, 2300 RA Leiden, the Netherlands \\
   $^{7}$  Max-Planck-Institute for Extraterrestrial Physics, Giessenbachstrae, 85748 Garching, Germany \\
   $^{8}$  Space Telescope European Coordinating Facility, European Southern Observatory, Karl-Schwarzschild-Str 2, 85748 Garching, Germany \\  
   $^{9}$  Gemini Observatory, Northern Operations Centre, 670 N. A'ohoku Place, Hilo, Hawaii 96720, USA \\
   $^{10}$ ASTRON - Netherlands Institute for Radio Astronomy, Postbus 2, 7990 AA Dwingeloo, The Netherlands\\
   $^{11}$ Kapteyn Astronomical Institute, University of Groningen Postbus 800, 9700 AV Groningen, The Netherlands\\
   $^{12}$ Max-Planck-Institute for Astrophysics, Karl-Schwarzschild-strasse 1, 85741 Garching, Germany \\
   $^{13}$ Universit\"ats-Sternwarte M\"unchen, Scheinerstr. 1, D-81679 M\"unchen, Germany \\
   $^{14}$ Centre for Astrophysics Research, University of Hertfordshire, Hatfield, Herts AL1 09AB, UK \\
   $^{15}$ Dunlap Institute for Astronomy \& Astrophysics, University of Toronto, 50 St. George Street, Toronto, Canada \\
  $^{16}$ Department of Physics, New Mexico Institute of Mining and Technology, Socorro, NM 87801, USA \\
}

\pagerange{\pageref{firstpage}--\pageref{lastpage}}
\pubyear{2011}

\def\aj{AJ}             
\def\araa{ARA\&A}       
\def\apj{ApJ}           
\def\apjl{ApJ}          
\def\apjs{ApJS}         
\def\aap{A\&A}          
\def\aaps{A\&AS}        
\def\mnras{MNRAS}       
\def\pasp{PASP}         

\begin{document}
\label{firstpage}

\maketitle

\clearpage

\begin{abstract}
We use the ATLAS$^{\rm 3D}$ sample of 260 early-type galaxies to study the apparent kinematic misalignment angle, $\Psi$, defined as the angle between the photometric and kinematic major axis. We find that 71\% of nearby early-type galaxies are strictly aligned systems ($\Psi \le 5\degr$), an additional 14\% have $5\degr < \Psi \le 10\degr$ and 90\% of galaxies have $\Psi \le 15\degr$. Taking into account measurement uncertainties, 90\% of galaxies can be considered aligned to better than $5\degr$, suggesting that only a small fraction of early-type galaxies ($\sim10$\%) are not consistent with axisymmetry within the projected half-light radius. We identify morphological features such as bars and rings (30\%), dust structures (16\%), blue nuclear colours (6\%) and evidence of interactions (8\%) visible on ATLAS$^{\rm 3D}$ galaxies. We use kinemetry to analyse the mean velocity maps and separate galaxies in two broad types of regular and non-regular rotators. We find 82\% of regular rotators and 17\% non-regular rotators, with 2 galaxies that we were not able to classify due to data quality. The non-regular rotators are typically found in dense regions and are massive. We characterise the specific features in the mean velocity and velocity dispersion maps. The majority of galaxies does not have any specific features, but we highlight here the frequency of the kinematically distinct cores (7\% of galaxies) and the aligned double peaks in the velocity dispersion maps (4\% of galaxies). We separate galaxies into 5 kinematic groups based on the kinemetric features, which are then used to interpret the ($\Psi - \epsilon$) diagram. Most of the galaxies that are misaligned have complex kinematics and are non-regular rotators. In addition, some show evidence of interaction and might not be in equillibrium, while some are barred. While the trends are weak, there is a tendency that large values of $\Psi$ are found in galaxies at intermediate environmental densities and among the most massive galaxies in the sample. Taking into account the kinematic alignment and the kinemetric analysis, the majority of early-type galaxies have velocity maps more similar to the spiral disks than to the remnants of equal mass mergers.  We suggest that the most common formation mechanism for early-type galaxies preserves the axisymmetry of the disk progenitors and their general kinematic properties. Less commonly, the formation process results in a triaxial galaxy with much lower net angular momentum.

\end{abstract}


%
%

\section{Introduction}
\label{s:intro}

The internal dynamics of early-type galaxies holds important clues about their formation. Looking at their morphological structure alone, early-type galaxies appear to be simple and uniform, but increasingly better observational technology and methods have revealed much more complex systems rich in internal dynamics and substructures.  Crucial for this were the kinematic observations of early-type galaxies \citep{1977ApJ...218L..43I, 1983ApJ...266...41D, 1983ApJ...266..516D, 1988A&A...193L...7B,1990A&A...239...97B} and two significant discoveries that some ellipticals rotate slowly \citep{1975ApJ...200..439B,1977ApJ...218L..43I} and that there are objects with significant rotation around the major axis \citep{1986ApJ...303L..45D,1988ApJS...68..409D,1989ApJ...344..613F,1989AJ.....98..147J}. This showed that among early-type galaxies there are systems with triaxial figures, perhaps even slowly tumbling \citep{1982MNRAS.198..303V,1982ApJ...263..599S}, and their internal structure is not determined by their total mass and angular momentum alone \citep[see the review by][]{1991ARA&A..29..239D}. 

Although discoveries of galaxies with rotation around the long axis were very exciting, the majority of galaxies seemed to show rotation around the apparent minor axis \citep{1979ApJ...229..472S, 1980MNRAS.193..931E, 1983ApJ...266...41D,1983ApJ...265..664D, 1988ApJS...68..409D,1988A&A...202L...5B,1989ApJ...344..613F,1989AJ.....98..147J,1990A&A...239...97B,1994MNRAS.269..785B}. Over the decade preceding the late-1990s, a picture emerged of elliptical galaxies exhibiting a range of properties with luminous objects having slow rotation, anisotropic velocity distributions, boxy isophotes and cores, taken to be indicative of triaxial figures, and less luminous galaxies having shapes flattened by rotation, isotropic velocity distributions, disky isophotes and cuspy cores, taken to be indicative of oblate figures \citep[for a synthesis see][]{1996ApJ...464L.119K}. 

The most straightforward evidence for triaxiality is the observation of a misalignment between the galaxy's angular momentum vector and the minor axis. In axisymmetric galaxies these two axes are aligned. Stationary triaxial shapes support four major types of regular stellar orbits: box orbits, short axis tubes, inner and outer long-axis tubes \citep{1985MNRAS.216..273D}. Given that among the tubes it is also possible to have both prograde and retrograde orbits, the combination of these major families will result in the total angular momentum vector pointing anywhere in the plane containing both the long and the short axis of the system \citep[e.g.][]{1987ApJ...321..113S}. Furthermore, it is also possible to have a radial variation of the relative weights assigned to different orbital families which will give rise to radially different kinematic structure and contribute to the radial variation of the observed misalignment \citep[see][for a detailed orbital analysis of a triaxial system]{2008MNRAS.385..647V}. 

In addition to the orbital origin of the misalignment between the shape of the system and its internal kinematics, it is also possible to observe a misalignment from pure projection effects, given that the orientation of a triaxial galaxy towards an observer is random. Hence, the angle at which the apparent minor axis of the observed (projected on the sky) galaxy is seen will be different from the angle of the projected short axis of the galaxy \citep{1956ZA.....39..126C,1977ApJ...213..368S,1979SvAL....5...37K}. When this is combined with a projection of the angular momentum vector, which depends on the specific orbital structure, we expect that the misalignment between the angular momentum vector and the principle axis will be observed regularly. This was beautifully illustrated by \citet{1991AJ....102..882S} with the montage of velocity maps of a triaxial model viewed at different projection angles. 

The combination of the apparent orientation of the total angular momentum, and the apparent shape of the system, can be used to statistically constrain the intrinsic shape of early-type galaxies as a family of objects, including the case when figure rotation is present \citep{1985MNRAS.212..767B}. The first analysis of the apparent misalignment angle, defined as $\tan \Psi = v_{min}/v_{maj}$, where $v_{min}$ and $v_{maj}$ are velocity amplitudes along the minor and major axis, was presented by \citet{1991ApJ...383..112F}. They compiled from the existing literature all galaxies for which it was possible to estimate $\Psi$ reasonably well and obtain their ellipticities. This compilation confirmed that a majority of early-type galaxies indeed had small misalignments,  with a few cases showing long-axis rotation (rotation around the major (long) axis). In terms of intrinsic shape of early-type galaxies, their results showed a wide range of acceptable solutions including distributions of only nearly oblate shapes, oblate and prolate shapes, as well as purely triaxial shapes.

The SAURON survey \citep{2002MNRAS.329..513D} provided velocity maps reaching to about one effective radius for a sample of nearby early-type galaxies. The survey confirmed the main findings of the previous decades and established that many of the dynamical properties of early-type galaxies are related to a measure of their specific angular momentum, which was available for the first time from velocity and velocity dispersion maps \citep{2004MNRAS.352..721E}. Based on their apparent angular momentum, the early-type galaxies separate into slow and fast rotators \citep{2007MNRAS.379..401E}, where slow rotators are weakly triaxial, but not far from isotropic, while fast rotators are nearly axisymmetric, intrinsically flatter and span a large range of anisotropies \citep{2007MNRAS.379..418C}.  Furthermore, the fast rotators are aligned, while slow rotators are misaligned \citep{2004MNRAS.352..721E,2007MNRAS.379..418C}. This global property is followed locally where fast rotators do not show radial changes in the orientation of the velocity maps, which is, however, typical for slow rotators  \citep{2008MNRAS.390...93K}.

In addition, the SAURON velocity maps of slow rotators exhibit a variety of kinematic structures, such as kinematic twists, kinematically distinct cores or showing no rotation at all, while the velocity maps of fast rotators are kinematically more uniform showing disk-like kinematics \citep{2008MNRAS.390...93K}. This suggests that the difference in the appearance of the velocity maps of these two types of galaxies is related to their internal structures and is a consequence of their (different) evolution paths. The features visible on the kinematic maps are the end products of various processes and it is potentially useful to assess their relative importance. 

The SAURON survey found 25\% of slow rotators among the nearby early-type galaxies. The SAURON sample, however, is not representative of the luminosity function of early-type galaxies and a question remains: what is the relative fraction of galaxies consistent with being axisymmetric? This question is relevant for our understanding of the importance of gas dissipation in the formation of early-type galaxies via hierarchical merging. Collisionless mergers of roughly equal mass progenitors generally produce triaxial galaxies, while gas dissipation generates nearly axisymmetric systems with disks \citep[e.g.][]{2006MNRAS.372..839N, 2007MNRAS.376..997J, 2009ApJ...705..920H}. Observations of molecular, atomic and ionised gas  \citep{2002AJ....123..729O,2006MNRAS.366.1151S, 2006MNRAS.371..157M, 2008ApJ...676..317Y,2008A&A...483...57S, 2009MNRAS.393.1255C,2010MNRAS.409..500O} suggest that the evolution of early-type galaxies is significantly influenced by gas reservoirs, both free or bound to other galactic systems. The presence of gas inevitably results in dissipation playing a major role in evolution.

The purpose of this work is three fold: (i) to analyse the kinematic maps and images of the volume limited sample of nearby early-type galaxies gathered by the ATLAS$^{\rm 3D}$ Survey \citep[][hereafter Paper I]{2011Cappellari}, (ii) to characterise quantitatively the morphological and kinematic features and determine their frequency, and (iii) to measure the kinematic misalignment angle exploiting the completeness of the sample and the two dimensional coverage of the kinematic data. Specifically, we explore the connection between the kinematic misalignment, the morphology and kinematic structures of nearby early-type galaxies. In this respect, this papers follows Paper I and its main results are used in \citep[][hereafter Paper III]{2011Emsellem} 

In Section~\ref{s:samp} we briefly describe the ATLAS$^{\rm 3D}$ sample and the types of data used in this paper. In Section~\ref{s:morph} we characterise the morphological and kinematical structures observed in the sample with more emphasis given to the latter. We separate early-type galaxies based on their rotation, define various features visible on the kinematic maps and identify galaxies according to these properties. This is followed by definitions of kinematic and photometric position angles, and the description of how these, as well as the ellipticity of the galaxies, were measured together with an estimate of the uncertainty (Section~\ref{s:detr}). The distribution of the kinematic misalignment angle is shown in Section~\ref{s:kinmis}, which is followed by a discussion (Section~\ref{s:discuss}) and conclusions  (Section~\ref{s:conc}).

%
%

\section{Sample and observations}
\label{s:samp}

The ATLAS$^{\rm 3D}$ sample and its selection is described in detail in Paper I. Here we briefly outline the main properties. Our galaxies were selected from a parent sample of objects brighter than $M_K < -21.5$ mag and a local volume with radius of $D = 42$ Mpc using the observability criterion that the objects have to be visible from the William Herschel Telescope (WHT) on La Palma: $|\delta - 29\degr| < 35 $, where $\delta$ is the sky declination, excluding the dusty region near the Galaxy equatorial plane. Galaxies were selected using the 2MASS extended source catalog \citep{2000AJ....119.2498J}, while the classification of  early-type galaxies was based on visual inspection of available imaging: SDSS and DSS colour images. Here the main selection criterion was the lack of spiral arms or dust lanes in highly inclined galaxies, following the Hubble classification \citep{1936RNeb..........H,1991trcb.book.....D} as outlined in \citet{1961hag..book.....S}.  The final sample contains 260 nearby early-type galaxies. 

Kinematic data used in this study were obtained using the SAURON integral-field spectrograph \citep[IFS;][]{2001MNRAS.326...23B} mounted on the WHT. SAURON is an IFS with a field-of-view (FoV) of $33 \times 41\arcsec$. The observing strategy and the data reduction is also described in detail in Paper I. The SAURON FoV, or mosaics of two SAURON pointings, was oriented along the major axis of the galaxies such as to maximise the coverage. Typically maps encompass one effective radius although for the largest galaxies only half of the effective radius is fully covered (see Paper III). The data reduction follows procedures described in \citet{2001MNRAS.326...23B} and \citet{2004MNRAS.352..721E}. For 212 galaxies we used publicly available SDSS Data Release 7 {\it r}-band images \citep{2009ApJS..182..543A}. For galaxies which were not observed by the SDSS we had imaging campaigns using the Wide Field Camera on the Isaac Newtown Telescope on La Palma, also in {\it r}-band. There we observed 46 galaxies and the data reduction and calibrations are presented in \citet{2011Scott}. Finally, there were two galaxies for which we were not able to obtain {\it r}-band images and in this study we used Two Micron All Sky Survey (2MASS) K-band observations instead.

%
%

\section{Characterisation of morphological and kinematic structures in the ATLAS$^{\rm 3D}$ Sample}
\label{s:morph}

In this section we describe morphological and kinematic features found in ATLAS$^{\rm 3D}$ galaxies. We are primarily interested in highlighting the existence of bars, rings, shells or other interaction features, as well as the existence of dusty disks or filamentary structures on the images. We also analyse the mean velocity maps and describe the kinematic features and their frequency in our sample. We point out the most significant features and accordingly sort galaxies in five kinematic groups which will be used in the rest of the paper. Additional remarks, images of velocity maps of the full sample and a table with the morphological and kinematic characteristics of galaxies  are presented in Appendices~\ref{A:cav},~\ref{A:maps} and~\ref{A:master}.

\subsection{Morphological features}
\label{ss:morph}

Our morphological characterisation is purely visual, based on SDSS and INT {\it r}--band images, as well as the SDSS true colour (red-green-blue) images \citep{2004PASP..116..133L} when available, but we do not attempt to quantify the amount of dust, the structure of shells or tidal tails, or the properties of bars. Our goal is to measure the frequency of obvious structures as they are visible on our {\it r}-band images. Occasionally, for confirmation of not clearly recognisable bars, we also use information contained in absorption and emission-line maps. A summary of morphological features found in ATLAS$^{\rm 3D}$ galaxies is given in Table~\ref{t:morph}, while the SDSS and INT colour images of the galaxies are shown in Paper I.  

\begin{table}
\caption{A summary of morphological features in ATLAS$^{\rm 3D}$. }
\label{t:morph}
\begin{tabular}{lcccc}
\hline
Feature & Number& Dust Disk& Filaments& Blue features\\
(1)& (2)&(3) &(4) &(5)\\
\hline
N               &159  &18 [16] & 8 [7]    &7 [4]\\
B               & 35   &2 [1] &  1  [0]   &2 [0]\\
R               & 13   & 3 [2] &  1 [0]   &2 [0]\\
BR            & 30   &1 [1]     &  3 [1]       &2 [0]\\
S                &9      &0 [0]    &  1 [0]     &1 [0]\\
I                 &12     &0 [0]     &  6 [5]      &1 [0]\\
\hline
\end{tabular}
\\
Notes: The total number of galaxies is 260; morphological and dust features were not classified in 2 galaxies without SDSS or INT imaging. 
Column (1): morphological features: {\it N} - no feature, regular shape. {\it B} - bar, {\it R} - ring,  {\it BR} - bar and ring,   {\it S} - shells, {\it I} - any other evidence for interaction.
Column (2): Number of galaxies with morphological features in Column (1).
Column (3): Galaxies with dust disks.
Column (4): Galaxies with dust filaments. 
Column (5): Galaxies with blue colour features.
Column (3-5): Within brackets is the number of galaxies that only have the features listed in that column (a dusty disk, dusty filaments or a blue feature)
\end{table}

Bars are detected in $\sim25\%$ of the galaxies in our sample (65 galaxies), while rings are seen in $\sim17\%$ of the sample (43 galaxies). Rings and bars often occur together, and about half of the barred systems have clearly visible rings, but there are 13 ringed systems with no obvious bar like structure. The rings in these systems resemble resonance rings (they do not appear as polar or collisional rings). There are three cases of dusty and blue, possibly star forming, rings (NGC3626, NGC4324 and NGC5582). The total fraction of galaxies with bars and/or rings increases to 30\% (78 galaxies). This is still likely a lower limit, but if we consider only galaxies with de Vaucouleurs type between -3 and 0 (175 galaxies in ATLAS$^{\rm 3D}$ sample), 45\% of galaxies have bars/rings in our sample.  This is in an excellent agreement with a recent near-infrared survey of barred S0 galaxies \citep{2009ApJ...692L..34L}.

We looked for dust using the same r-band SDSS and INT images. We found 24 systems with dust in ordered disks, and 20 systems with filamentary dusty features, giving the total fraction of dusty systems of 18\%.  An inspection of colour images reveals 15 galaxies (6\%) with some evidence of blue colours, half of which are found in the nuclei and half in (circum-nuclear) rings. Here we report the obvious cases, and their number is likely a lower limit only, but this does not influence the results of the paper. Note there are some well known cases of nuclear dust disks visible from space-based observations \citep[e.g. NGC4261,][]{1996ApJ...460..214J}, which we do not see on our ground-based images. We do not include them in our statistics. Similarly, we do not look for other morphological features below the spatial resolution of our images (e.g. nuclear bars). 

Evidence for past interaction of various degrees are seen in 21 (8\%) galaxies (based on our images from the SDSS and INT). These objects are likely at different stages of interaction, but they are mostly not actively merging systems. In particular, shells are visible in 9 systems at our limiting surface brightness of $~\sim26$ mag/arcsec$^2$. Evidence of past interactions are visible at all environmental densities, but they do not  occur in galaxies which have other morphological perturbations (such as bars) at our surface brightness limit. In most cases, the interacting galaxies do not show ordered dusty disks in the central regions. Filamentary dust features, however, can be found in half of the interacting galaxies. A specific study of shells and other interaction features based on deeper MegaCam images will be a topic of a future paper in the series.

\subsection{Kinematic structures}
\label{ss:kin}

The majority of velocity maps of early-type galaxies in our sample show ordered rotation. More complex features, although present, are not common.  We use the mean velocity and the velocity dispersion maps to perform a complete description of kinematic structures that occur in the early-type galaxies of our volume limited sample. 

\subsubsection{Two types of rotation}
\label{sss:class}

We performed an analysis similar to  \citet{2008MNRAS.390...93K} using kinemetry\footnote{The IDL \textsc{kinemetry} routine can be found on http://www.eso.org/$\sim$dkrajnov/idl.} \citep{2006MNRAS.366..787K} on velocity maps. This method consists of finding the best fitting ellipse along which the velocities can be described as a function of a cosine change in the eccentric anomaly. In that respect kinemetry is a generalisation of isophotometry of surface brightness images \citep{1978MNRAS.182..797C,1985ApJS...57..473L,1987A&A...177...71B,1987MNRAS.226..747J} to other moments of the line-of-sight velocity distribution (the mean velocity, velocity dispersion, etc.). This means that the stellar motions along this ellipse can be parametrized by a simple law, $V=V_{rot} \cos(\theta)$, where $V_{rot}$ is the amplitude of rotation and $\theta$ is the eccentric anomaly. Note that the same expression describes the motion of gas clouds on circular orbits in a thin (inclined) disk \citep[e.g.][]{1997MNRAS.292..349S,2004ApJ...605..183W} and that, when kinemetry is applied to the velocity maps of thin gas disks, it achieves similar results to the titled-ring method \citep[e.g.][]{1987PhDT.......199B, 1990ApJ...364...23S,1994ApJ...436..642F}. There are, however, conceptual differences. The tilted-ring method determines the best-fitting ellipse by fitting a cosine function in a least-squares sense along an elliptical path. Instead, kinemetry performs a rigorous generalisation of the photometric ellipse fitting. It determines the best-fitting ellipse by minimising the Fourier coefficients up to the 3rd, except the  $\cos(\theta)$ term. This ensures a more robust fit and ensures that the higher order Fourier terms are unaffected by the ellipse fit. Moreover, the approach adopted by kinemetry allows the same method to be used to fit both photometric and kinematic data. The method first fits for the ellipse parameters, position angle $\Gamma_{kin}$ and flattening of the ellipse $q_{kin}$. The velocity profile along the best fitting ellipse is then decomposed into odd Fourier harmonics. The first order $k_1$ is equivalent to $V_{rot}$, while the higher order terms show departures of the velocity profiles from the assumed cosine law. Examples of typical kinemetric radial profiles of the 48 early-type galaxies from the SAURON survey, most of which are also part of ATLAS$^{\rm 3D}$ sample, can be found in Appendix B of \citet{2008MNRAS.390...93K}, while examples of residual velocity maps obtained subtracting kinemetry fits are shown in \citet{2006MNRAS.366..787K}.

Deviations from the cosine law can be quantified by measuring the amplitude of the $k_5$ harmonics. In practice it is better to use a scale free measure which is given by dividing $k_5$ with local rotation $k_1$. In order to characterise each object, we use the radial profiles to calculate the luminosity weighted average ratio $\overline{k_5/k_1}$, following the prescription from \citet{1999ApJ...517..650R}. We exclude rings for which kinemetry was not able to find a good fit (i.e. the ellipse flattening hits the boundary value).  We estimate the uncertainty on $\overline{k_5/k_1}$ with a Monte Carlo approach by perturbing each point of the $k_5/k_1$ radial profile based on its measurement error, calculate the luminosity-weighted average and repeat the process 1000 times. The uncertainty is the standard deviation of the Monte Carlo realisations. The values of  $\overline{k_5/k_1}$ are determined within one effective radius or within the semi-major axis radius of the largest best fitting ellipse that is enclosed by the velocity map. 

We set a limit of $\overline{k_5/k_1} < 0.04$ for the velocity map to be well described by the cosine law. The choice for this number is somewhat arbitrary, but we based it on the mean uncertainty on $k_5/k_1$ for all galaxies ($\sim 0.03$) and the resistant estimate of its dispersion ($\sim 0.01$). Note that this is higher than  the 2\% used by \citet{2008MNRAS.390...93K}, but the observations of the SAURON sample were of higher signal-to-noise ratio and lower average uncertainty on $k_5/k_1$ (0.015). If $\overline{k_5/k_1}$ is larger than 4\%, we flag the velocity map as not being consistent with the cosine law. In this way we separate two types of rotations among early-type galaxies. 

Galaxies of the first type, consistent with having $\overline{k_5/k_1}<0.04$, have velocity maps dominated by ordered rotation. These we call  {\it Regular Rotators} (RR).  Galaxies of the second type, consistent with $\overline{k_5/k_1}>0.04$ have velocity maps characterised by more complex structures, including cases where rotation is not detectable. As a contrast to the RR galaxies we call them {\it Non-Regular Rotators} (NRR) galaxies.  The majority of objects in the ATLAS$^{\rm 3D}$ sample belong to the RR type (214 or 82\%), while there are 44 (17\%) objects of the NRR type. We were not able to classify two galaxies (PGC058114 and PGC170172) due to the low signal to noise ratio and an unfortunate position of a bright star.

\begin{figure*}
        \includegraphics[width=\textwidth]{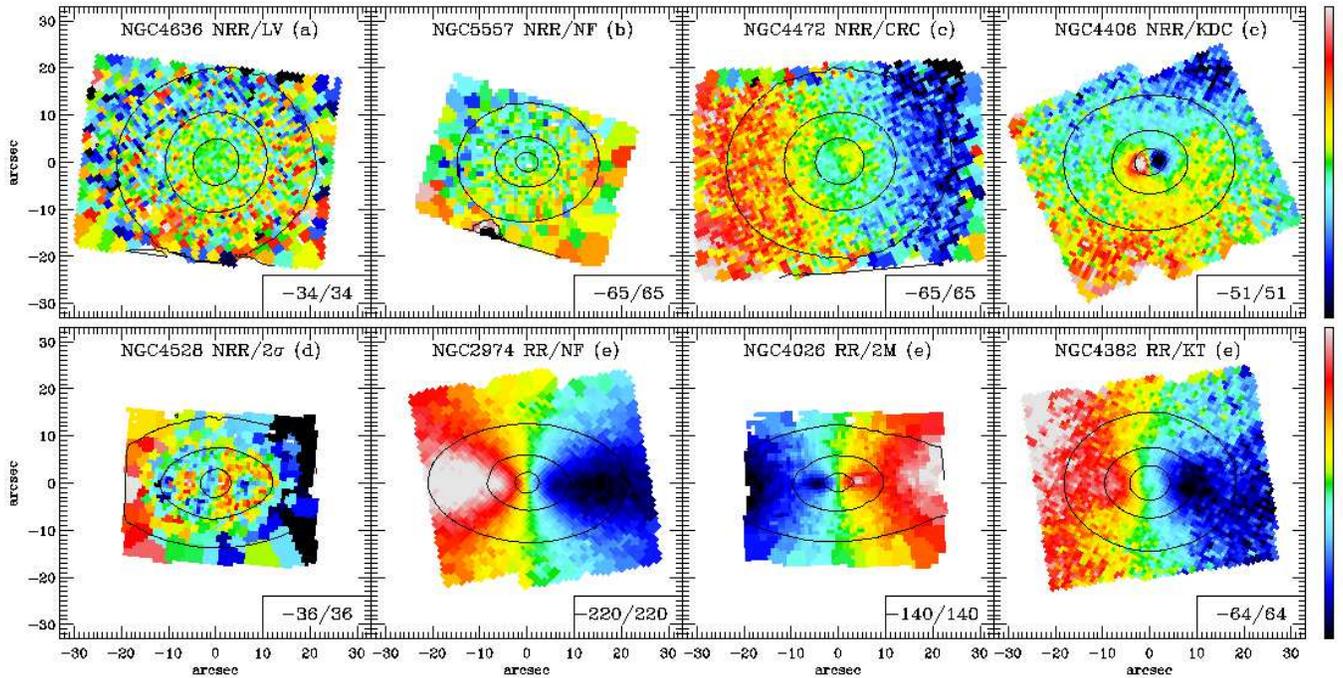}
\caption{\label{f:velmaps} Example of various features found on the mean velocity maps of ATLAS$^{\rm 3D}$ galaxies. From left to right, top to bottom: NRR/LV (NGC4636) part of group {\it a}, NRR/NF (NGC5557) part of group {\it b}, NRR/CRC (NGC4472) and NRR/KDC (NGC4406) part of group {\it c}, NRR/2$\sigma$ (NGC4528) part of group {\it d}, and representing group {\it e} RR/NF (NGC2974), RR/2M (NGC4026) and RR/KT (NGC4382). All maps are oriented such that the large scale photometric major axis is horizontal. Values in the lower right corners show the range of the plotted velocities in km/s.  For definition of kinematic groups see Tables~\ref{t:groups}.}
\end{figure*}

\subsubsection{Kinemetric features}
\label{sss:features}

The majority of the velocity maps are dominated by ordered rotation, but there are several distinct features recognisable on the kinematic maps, especially among the galaxies of the NRR type. The diversity of the kinematic features suggest a variety of formation processes at work in early-type galaxies. We wish to describe these fossil records and quantify their frequency among the two rotation types. As above, we use the kinemetry analysis within one effective radius (or within the semi-major axis radius of the largest best fitting ellipse that is enclosed by the velocity map) to define various kinematic features occurring in our sample:

\begin{itemize}

\item {\it No Feature} (NF) velocity maps are flagged if the orientation of the best fitting ellipses, $\Gamma_{kin}$, is constant with radius (both for RR and NRR type of rotation).  In the case of NRR galaxies with a measurable rotation, $\Gamma_{kin}$ can also change erratically between adjacent rings.

\item {\it Double Maxima} (2M) have radial profiles of the $k_1$ parameter characterised by a rapid rise of the velocity reaching a maximum value, which is followed by a decrease and subsequent additional rise to a usually larger velocity. These velocity maxima are aligned.

\item {\it Kinematic Twist} (KT) is defined as a smooth variation of $\Gamma_{kin}$ with an amplitude of at least $10\degr$ over the map.

\item {\it Kinematically Distinct Core}\footnote{There is some confusion in the literature on the naming of these kinematic structures. Both Decoupled/Distinct and Core/Component terms are used to specify the same thing. We choose to use the combination of Distinct Cores in order to stress that they happen in the central regions of the galaxies but they might not be dynamically decoupled from the rest of the system.} (KDC) is defined when there is an abrupt change in $\Gamma_{kin}$ with a difference larger than $30\degr$ between adjacent components, and $k_1$ drops to zero in the transition region. We require that at least two consecutive rings have a similar $\Gamma_{kin}$ measurement within the component. 

\item {\it Counter-Rotating Core} (CRC) is a special case of KDC where the change in $\Gamma_{kin}$ is of the order of $180\degr$. 

\item {\it Low-level Velocity} (LV) map is defined when $k_1 < 5$ km/s. In these cases rotation is not measurable and kinemetry cannot determine the ellipse parameters. 

\item {\it Double $\sigma$} (2$\sigma$) are found by visually inspecting the velocity dispersion maps. This feature is characterised by two off centre, but symmetric peaks in the velocity dispersion, which lie on the major axis of the galaxy. We require that the distance between the peaks on the velocity dispersion map is at least half the effective radius. 

\end{itemize}

\begin{table}
\caption{A summary of kinemetric types and features in the ATLAS$^{\rm 3D}$ sample. }
\label{t:class}
\begin{tabular}{ccccl}
\hline
Feature & RR & NRR & Comment\\
\hline
NF                  &171 &  12 & No Feature on the map\\
2M                  & 36  & 0    & Double Maxima in radial velocity profile\\
KT                   & 2    &  0  &  Kinematic Twist\\
KDC               & 0    &  11&  Kinematically Distinct Core\\
CRC               & 1   & 7   &   Counter-Rotating Core\\
2$\sigma$     &4    &  7  &   Doube peak on $\sigma$ map\\
LV                   &0    &  7  &   Low-Level velocity (non rotator)\\
\hline
\end{tabular}
\\
Notes: The total number of galaxies is 260 and two galaxies were left unclassified in terms of their kinematic features. 
 
\end{table}

In principle, each map could be characterised by a combination of a few of the above features.  Specifically, all features could occur both in RR and NRR galaxies, but this is generally not the case, as it can be seen in Table~\ref{t:class}, which summarises the kinemetric features found in the ATLAS$^{\rm 3D}$ sample. This can be understood by considering that any feature on the velocity map, especially those associated with the change of $\Gamma_{kin}$, will disturb the map such that $k_5/k_1$ will increase. Unless those features are small (relative to 1 R$_e$), the $\overline{k_5/k_1}$ will be larger than 4\% and, hence, the galaxy will be classified as NRR. An exception is the 2M feature since the two maxima are aligned and the $k_5/k_1$ might increase only within the region of the rotation dip \citep[see][for examples of the model velocity maps and their analysis]{2006MNRAS.366..787K}. For a discussion on differences between 2M, KDC, $2\sigma$ and CRC galaxies see Appendix~\ref{A:rem}.

As stated above, the majority of galaxies are of RR type and they do not have any specific feature (66\%). The second most common feature  (14\%) are the aligned maxima on the velocity maps (2M) and they occur only in the RR type. In a few cases, RR type galaxies are also found to show KT (2), CRC (1) and 2$\sigma$ (3) features. The mean velocity maps of these more complex kinematic features are typically only marginally consistent with being of RR type. 

The number of galaxies with different features are evenly spread among the NRR type. There are 18 (7\%) galaxies that have a KDC or a CRC feature (11 and 7 objects respectively), 12 ($\sim5$\%) do not show any features, 7 ($\sim3$\%) do not have any detectable rotation, while 7 have 2$\sigma$ peaks on velocity dispersion maps. It is possible that about a third of NRR/NF maps would be classified RR/NF in lower noise velocity maps. Possible candidates include: NGC770, NGC4690, NGC5500, and NGC5576. In Fig.~\ref{f:velmaps} we show examples of the velocity maps dominated by the typical kinemetric features.

\begin{figure*}
        \includegraphics[width=\textwidth]{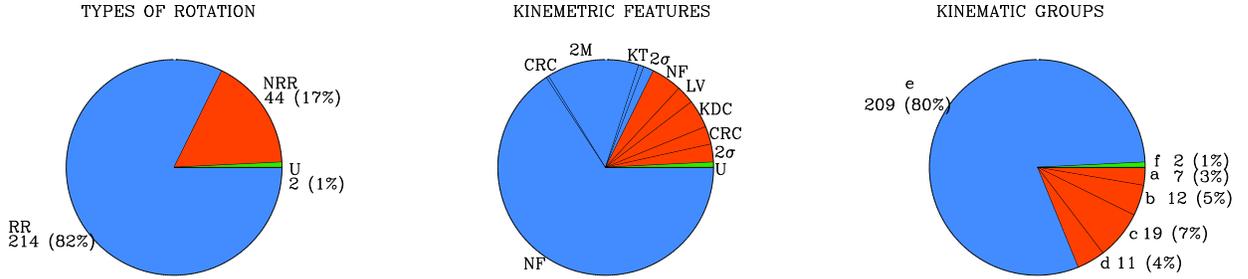}
\caption{\label{f:piecharts} Diagrams presenting the kinematic analysis of ATLAS$^{\rm 3D}$ galaxies. {\bf Left.} The frequency of two types of rotatiors: Regular Rotators (RR) and Non-Regular Rotators (NRR). {\bf Middle.} The kinemetric features. Only those features found in our sample are shown. The numbers of galaxies are not shown for clarity. They are given in Table~\ref{t:class}. {\bf Right.} The five kinematic groups comprising significant kinemetric features. Letters {\it a-f} are explained in Table~\ref{t:groups}.  In all diagrams blue colour refers to ordered velocity maps that can be described by the cosine law ($\overline{k_5/k_1} \leqslant 0.04$) and red colour to the complex velocity maps poorly described by the cosine law ($\overline{k_5/k_1} > 0.04$). Objects which were not classified are represented by the green slice and marked with 'U'. }
\end{figure*}
\subsubsection{Five kinematic groups of early-type galaxies}
\label{sss:5types}

In Section~\ref{sss:class} we quantified two types of rotation present in early-type galaxies, while in Section~\ref{sss:features} we discussed all features visible on kinematic maps in our sample. In Paper III we separate galaxies according to their specific (projected) angular momentum into fast and slow rotators. That separation is somewhat arbitrary and we use the two types of rotations on the velocity maps (RR and NRR) to empirically divide slow and fast rotators. In the rest of the paper we will continue to use the terminology of RR and NRR type of rotation, instead of Fast and Slow Rotators, but we emphasise the respective similarity between these definitions, although it is not {\it a priori} necessary that all RR galaxies are FR (see Paper III).

The majority of galaxies are {\it Regular Rotators} without specific features, while a minority of galaxies show a variety of kinematic substructures. Given the fact that certain features do not occur in one of the two types of rotation presents a constraint on galaxy formation.  In order to facilitate the usefulness of these features, we propose a system of five groups, which is based on reduction of non occurring features and blending of features with likely similar origin. 

In Table~\ref{t:groups} we summarise the five groups. Note that in Table~\ref{t:groups} we used only features which occur in our sample, but the intention is that {\it group a} consists of galaxies which do not show any rotation, while {\it group b} consists of galaxies with complex velocity maps, but which do not show any specific feature. {\it Group c} comprises kinematically distinct cores, including the sub-group of counter-rotating cores, while {\it group d} has galaxies with double peaks on the velocity dispersion maps. The most numerous is the {\it group e}, consisting of galaxies with simple rotation, and of galaxies with two aligned velocity maxima or with minor kinematic twists. In Fig.~\ref{f:velmaps} we link the typical kinematic features with the five significant kinematic groups.  

\begin{table}
\caption{Kinematic groups. }
\label{t:groups}
\begin{tabular}{ccl}
\hline
Group  & \# of galaxies & Feature \\
\hline
a       &  7      & NRR/LV                                                             \\
b       &  12    & NRR/NF                                                             \\
c       &  19    & NRR/KDC, NRR/CRC, RR/CRC                   \\
d       &  11    & NRR/$2\sigma$, RR/$2\sigma$                    \\
e       &  209 & RR/NF, RR/2M, RR/KT                                     \\
f         &  2     & U                                                                          \\
\hline
\end{tabular}
\\
Notes: The last row is reserved for galaxies for which we were not able to determine kinematic features and which remain {\it Unclassified}.  
\end{table}

\begin{figure}
        \includegraphics[width=\columnwidth]{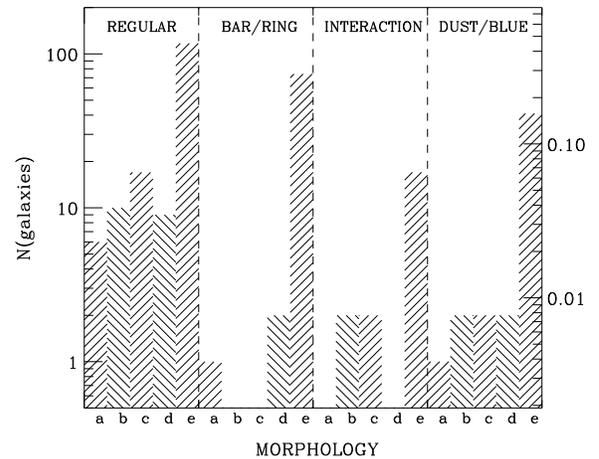}
\caption{\label{f:photkin} Histogram with the comparison of morphological and kinematic features in ATLAS$^{\rm 3D}$ galaxies. The morphological features are binned into {\it Regular} (featureless and regular shapes), {\it Bar/Ring} (bars and/or rings),  {\it Interaction} (shells or other interaction features) and {\it Dust/Blue} (dusty filaments, dusty disks or blue nuclear features). Hatched vertical bars show the number of galaxies having that morphological feature and being part of one of the five kinematic groups (a,b,c,d,e). The right hand axis is in the units of total number of galaxies in the sample. }
\end{figure}

The three pie-chart diagrams in Fig.~\ref{f:piecharts} visualise the frequency of the two types of rotation, different kinemetric features and their inclusion to significant kinematic groups. As mentioned before, the majority of early-type galaxies in the local universe are ordered, {\it Regular Rotators}. There are, however, a number of different kinemetric features visible on the maps of the mean velocity and velocity dispersion, but they mostly occur in {\it Non-Regular Rotators}. Finally, the last diagram shows the relative frequency of the five most significant kinematic groups in the ATLAS$^{\rm 3D}$ sample.  Note that the number of {\it e} galaxies  (209) is not equal to the number of RR systems (214). The reason is that  that 1 RR/CRC and 4 RR/$2\sigma$ galaxies were put together with other NRR/CRC and NRR/$2\sigma$ systems into groups {\it c} and {\it d}. 

For a discussion on possible caveats of the kinemetric analysis we refer the reader to Appendix~\ref{A:cav}, while in Appendix~\ref{A:maps} we show the velocity maps of all ATLAS$^{\rm 3D}$ galaxies sorted in their kinematic groups.

\begin{figure*}
        \includegraphics[width=\textwidth]{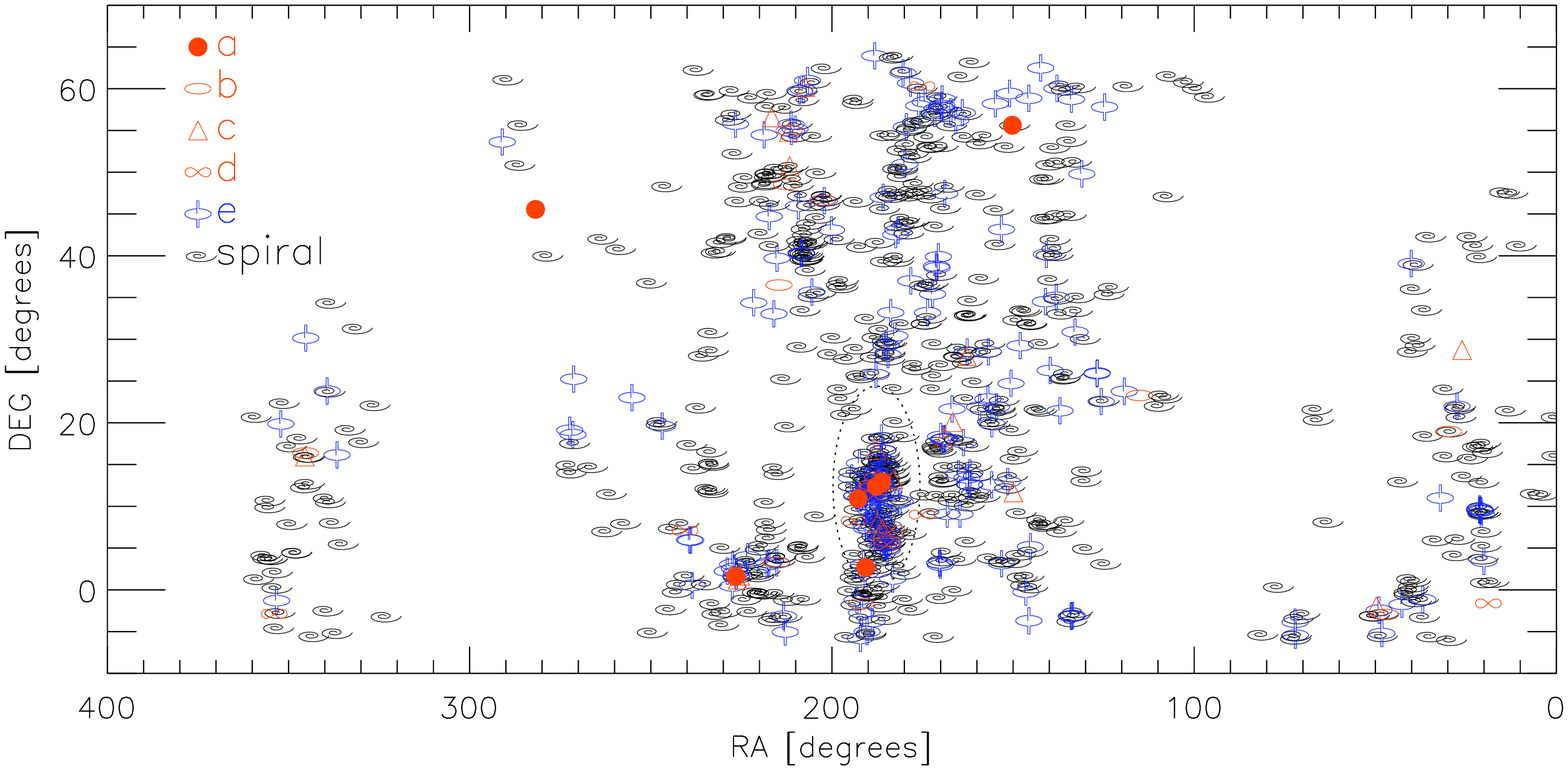}
        \includegraphics[width=\textwidth]{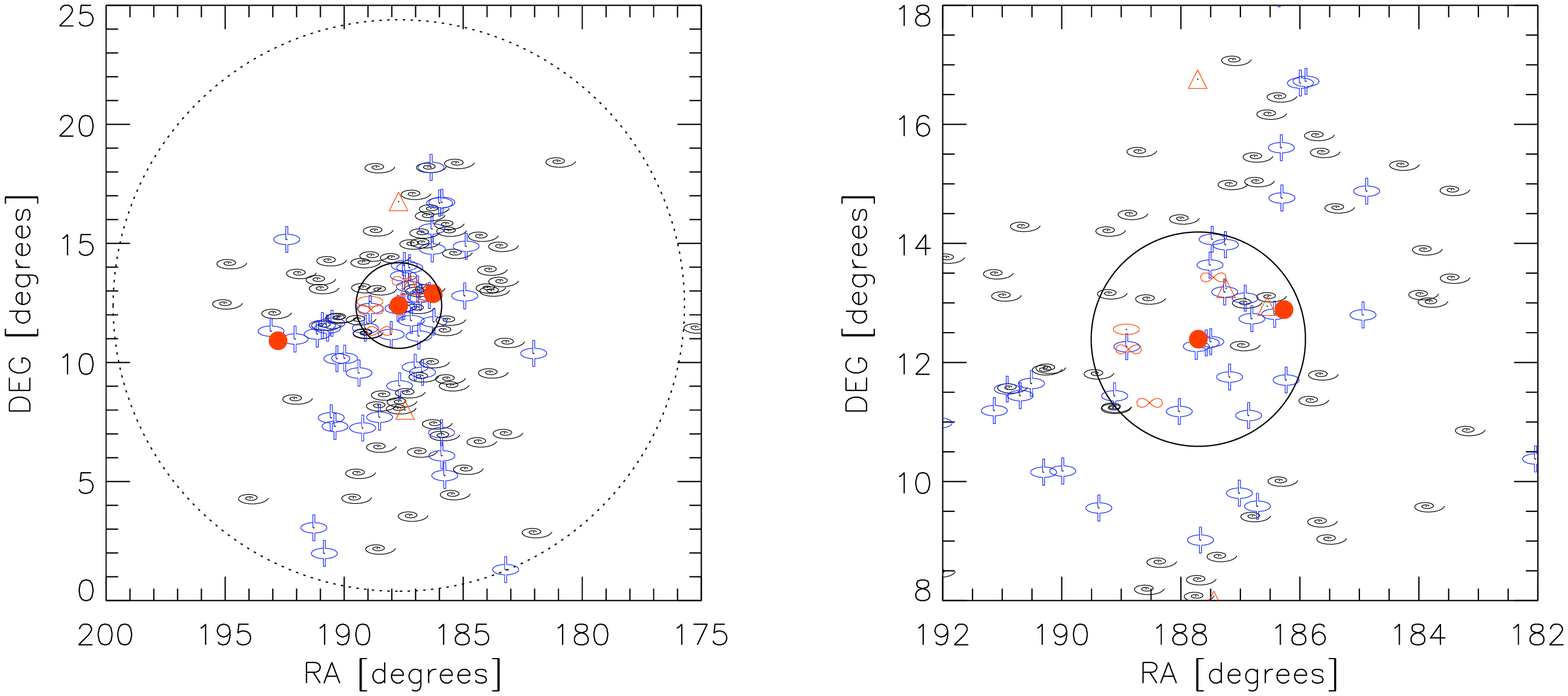}
\caption{\label{f:envkin} The spatial distribution of kinematic groups of ATLAS$^{\rm 3D}$ galaxies. All galaxies with M$_{K} < -21.5$ mag (the parent sample) are shown: full spatial distribution {\bf (top)}, the Virgo cluster galaxies {\bf (bottom left)} and a zoom in on the core of the Virgo cluster {\bf (bottom right)}. The legend on the top panel describes the symbols on all plots showing galaxies in kinematic groups: {\it a} (no rotation), {\it b} (Non-Regular Rotators without special kinematic features), {\it c} (kinematically distinct cores, including counter-rotation cores),  {\it d} ($2\sigma$ peak galaxies), and {\it e} (Regular Rotators). Spiral galaxies are shown as logarithmic spirals. The large dotted circles on bottom left panel has a radius of 12\degr and encompasses the same region as the dotted ellipse in the top panel. The Virgo cluster core is shown by the solid circle centred on M87 with R = 0.5 Mpc.  }
\end{figure*}

\subsubsection{Linking morphology, kinematics and environment}
\label{sss:links}
 
Figure~\ref{f:photkin} shows a histogram of morphological features of ATLAS$^{\rm 3D}$ galaxies. We created four bins grouping objects with resonance phenomena (including bars, rings and bars with rings), interaction features (including shells and other interaction characteristics), dust/blue (including filamentary dust, dust disks and the blue nuclear colours) and featureless galaxies with regular early-type morphology. In the same histogram we added the frequency of galaxies of the five kinematic groups for a given morphological feature.  It is hardly surprising that in all morphological bins the most represented are the galaxies of the group {\it e} (RR galaxies), given that this is also the most numerous group. 

It is somewhat surprising that galaxies with evidence for interaction do not show more complex kinematics (only two galaxies are from groups {\it b} and {\it c}), which is probably due to the difference in the dynamical state and time scales between the large (interaction features) and small scales (kinematics). The resonance phenomena are linked to disk dominated systems, and almost all galaxies in this bin show Regular Velocities. There are three exceptions of which one deserves special attention:  a barred NRR/LV (NGC4733), which is seen at very low inclination and, hence, likely an object with intrinsic rotation. Dust or blue nuclear features are also present in galaxies of all groups with complex kinematics, but only in one or two galaxies per group. In this group there is also a special case: a round NRR/LV galaxy (NGC3073) which also has blue UV colours \citep{2007ApJS..173..597D}.

Complex kinemetric features (groups {\it a, b, c} and {\it d}) are mostly found in galaxies with typical, featureless, early-type morphologies, confirming the reputation of early-types that while looking simple, they retain complex internal structure. All intrinsically non-rotating galaxies (group {\it a}) are here, as well as the majority of galaxies with KDCs or just complex velocity maps. Galaxies with two peaks on the velocity dispersion maps are also mostly found in this bin. This suggests that any process that shaped these galaxies has happened a long time ago.

As an illustration of the environmental influence on the kinematics of galaxies (and the membership to a specific kinematic group) we show in Fig.~\ref{f:envkin} the distribution on the sky of all galaxies brighter than -21.5 mag in K band of the parent sample (Paper I; both ATLAS$^{\rm 3D}$ and spiral galaxies). The kinematic groups are distinguished by different symbols. The top panel shows the northern hemisphere and, excluding the Virgo cluster, it can be seen that the spirals and galaxies from the group {\it e} have a relatively similar spatial distribution, while the galaxies with complex kinematics are found typically surrounded, in projection, by other galaxies. Obvious exemptions are two {\it a} group galaxies in the Northern part of the plot: NGC3073 (RA $\sim150\degr$) and NGC6703 (RA $\sim280\degr$). Three other galaxies are in less densely populated regions: NGC5557 from group {\it b} (RA $\sim210\degr$), NGC661 from group {\it c} (RA $\sim 25\degr$) and NGC448 from group {\it d} (RA $\sim 20\degr$). For NGC3073 and NGC6703 there is evidence they are actually disks seen at low inclinations \citep{1991trcb.book.....D,2011Emsellem}, while both NGC448 and NGC661 have two $\sigma$ peaks in their velocity dispersion maps, but the separation between the peaks for NGC661 is below the imposed limit of 0.5 R$_{e}$, making it a very small feature (see Appendix~\ref{A:cav}). 

In the Virgo cluster the situation is alike: galaxies with RR type rotation and spirals are similarly distributed, although spirals tend to be further from the centre of the cluster. In contrast,  the galaxies with complex kinematics are mostly found in the very core of the cluster (R $ < 0.5$ Mpc). Specifically, there are 11 galaxies from groups {\it a}, {\it b}, {\it c} and {\it d} and 8 of them are within 0.5 Mpc in radius centred on the M87. The three outside are: NGC4489 (North of the core), NGC4472 (South of the core) and NGC4733 (East of the core). NGC4472 and NGC4489 are both classified as CRC galaxies, but NGC4472 (or M49) is in a more densely populated environment and it is the most massive galaxy of a small sub-group, while NGC4489 is in a region with a fewer larger galaxies. NGC4733, also in a less densely populated region, was mentioned above as a barred galaxy.

A version of Fig.~\ref{f:envkin} showing the fast/slow rotators instead of our five kinematic classes is presented in \citet[][hereafter Paper VII]{2011bCappellari}. We refer to that paper for a detailed investigation of the connection between environment and kinematics of ETGs. In Paper VII, we find a clear excess of slow rotators in the densest core of the Virgo cluster. The distribution of galaxies with complex kinematics we find here (groups {\it a}, {\it b}, {\it c} and {\it d}) confirms that the environmental effects on the internal dynamics are significant.

%
%
\section{Determination of kinematic and photometric position angles and ellipticities}
\label{s:detr}

In this section we present methods for determining global values of the kinematic and photometric position angles as well as the global ellipticity of galaxies.  All values are tabulated in Table~\ref{t:master}.

\subsection{Kinematic Position Angle}
\label{ss:kinPA}

The global kinematic position angle (PA$_{kin}$) is the angle which describes the orientation of the mean stellar motion on a velocity map. It is usually defined as the angle between the north and the receding part of the velocity map (maximum values). If figure rotation is absent, PA$_{kin}$ is also perpendicular to the orientation of the apparent angular moment \citet{1988FranxThesis}. We measure it using the method outlined in Appendix C of \citet{2006MNRAS.366..787K}\footnote{We use an IDL routine  \textsc{fit\_kinematic\_pa.pro} publicly available on http://www.purl.org/cappellari/idl.}. Briefly, for any chosen PA$_{kin}$ we construct a bi-(anti)symmetric velocity map mirrored around an axis with the position angle PA$_{kin}+90\degr$. The best PA$_{kin}$ is defined as the angle which minimises the difference between the symmetrised and the observed velocity maps. 

The error on PA$_{\rm kin}$ is defined as the smallest opening angle that encloses the position angles of all the models for which the symmetrised and observed data are consistent within a chosen confidence level. The acceptable confidence level was defined by $\Delta\chi^2<9+3\sqrt{2N}$, where $\Delta\chi^2<9$ is the standard $3\sigma$ level for one parameter, and we included an additional term $3\sqrt{2N}$ to account for the $3\sigma$ uncertainties in $\chi^2$. The latter term becomes important when dealing with large datasets, as pointed out in a similar context by \citet{2009MNRAS.398.1117V}.

We produce 361 different bi-symmetrised maps with $0.5\degr$ steps in position angle ranging from 0 to $180\degr$. The actual uncertainty depends also on bin sizes, FoV, asymmetric coverage of the galaxy and on velocity errors (see Section~\ref{ss:errors}). In addition we compare and verify our results with the radial profiles of $\Gamma_{kin}$\footnote{Note that we differentiate between the global and local kinematic orientation, PA$_{kin}$ and $\Gamma_{kin}$ respectively, estimated with different methods. The same applies for the photometric values.} derived using kinemetry (see Section~\ref{s:morph}).  As was also shown in \citet{2006MNRAS.366..787K}, the average luminosity weighted $\overline{\Gamma_{kin}}$ obtained from kinemetry agrees well with the global PA$_{kin}$ for a typical velocity map.

\subsection{Photometric Position angle and ellipticity}
\label{ss:photPA}

The photometric position angle (PA$_{phot}$) measures the orientation of the stellar distribution and it defines the position of the apparent photometric major axis measured east of north. We derive PA$_{phot}$ by calculating the moments of inertia of the surface brightness distribution from the SDSS and INT {\it r}-band images. At the same time the method provides the global ellipticity $\epsilon$. The PA$_{phot}$  and $\epsilon$ estimated in this way are dominated by large scales. This is favourable since we want to derive the orientation and the shape representative of the global stellar distribution, particularly to avoid the influence of the bars, which are usually restricted to small radii and are common in our sample. For this reason, we also try to use the largest possible scales of the images.

We first determine the median level and the root-mean-square (rms) variation of the sky in each image. We then use an IDL routine that measures the moment of inertia\footnote{The IDL routine is called \textsc{find\_galaxy.pro} and is a part of the MGE package \citep{2002MNRAS.333..400C} that can be found on: http://www.purl.org/cappellari/idl.} on pixels that are a few times the sky rms above the zero (a median sky level was subtracted from the images). We masked the bright stars and companion galaxies if present. As levels we use 0.5, 1, 3 and 6 times the sky rms. The standard deviation of the measurements at these levels is used to estimate the uncertainties to PA$_{phot}$ and $\epsilon$. The final $\epsilon$ and PA$_{phot}$ are taken from the measurement obtained using pixels that were 3 times the rms. In some cases most of the galaxy surface brightness is dominated by the bar, and in order to probe the underlying disk one has to encompass the faint outer regions. In these cases, depending on the size of the bar, we use the measurements obtained at lower sky cuts, 0.5 or 1 times the sky rms. In this way, PA$_{phot}$ was typically measured between 2.5 to 3 effective radii. 

\begin{figure}
        \includegraphics[width=\columnwidth]{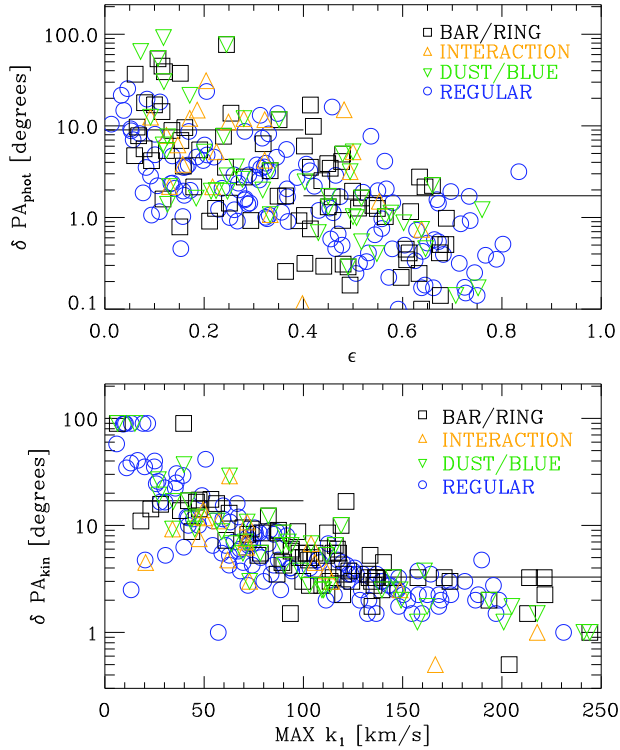}
\caption{\label{f:errorPA} {\bf Top:} The uncertainty on the photometric position angle, $\delta$PA$_{phot}$, as a function of ellipticity, {\bf Bottom:} The uncertainty on the kinematic position angle, $\delta$PA$_{kin}$, as a function of the maximum rotational velocity reached within the SAURON FoV. Horizontal lines on both panels show mean uncertainty values for the region they cover. On the top panel: $\sim9\degr$ for $\epsilon<0.4$ and $\sim 2\degr$ for $\epsilon>0.4$. On the bottom panel: $\sim17\degr$ for $k^{max}_1>100$ km/s and $\sim 3\degr$ for $k^{max}_1>100$ km/s. Black squares show galaxies with bars and/or rings, orange upward pointing triangles galaxies with evidences for interactions, green downward pointing triangles galaxies with evidence for dust or blue nuclei, and other, regularly looking early-type galaxies are shown with blue circles. }
\end{figure}
\begin{figure}
         \includegraphics[width=\columnwidth]{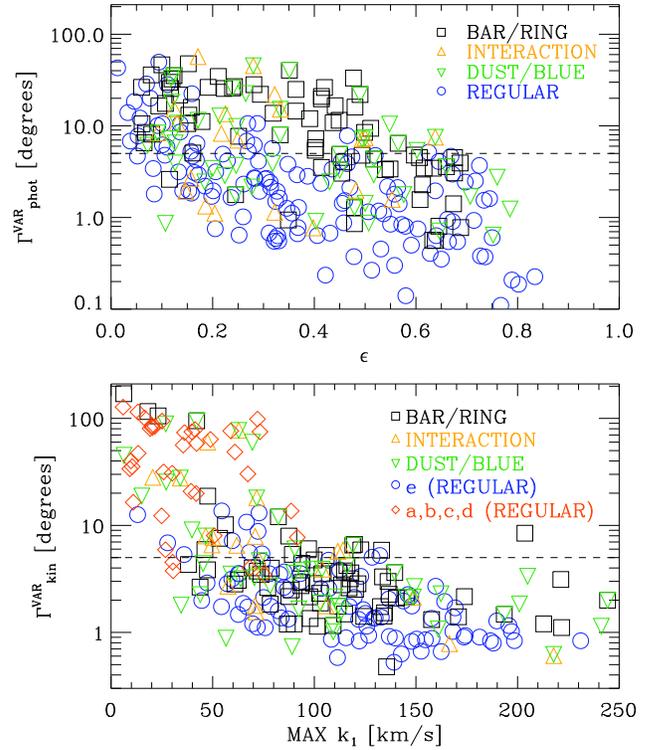}
\caption{\label{f:PAcompre} {\bf Top:} An estimate of the photometric radial variation of $\Gamma^{\rm VAR}_{phot}$ plotted as a function of ellipticity. {\bf Bottom:} An estimate of the kinematic radial variation of $\Gamma^{\rm VAR}_{kin}$ plotted as a function of the maximal rotational velocity within the SAURON FoV  {\bf (bottom)}. Dashed horizontal lines in both plots are at $5\degr$. Black squares are galaxies with bars and/or rings. Orange upward pointing triangles are interacting systems. Green downward pointing triangles are galaxies with evidence for dust or blue nuclei. On the top panel, regularly looking early-type galaxies are shown with blue circles. On the bottom panel, the red diamonds show galaxies from {\it a}, {\it b}, {\it c} and {\it d} kinematic groups with regular morphology. Galaxies from the group {\it e} with regular morphologies are shown with blue circles.}
\end{figure}

We also fitted ellipses to the isophotes of our galaxies and obtained radial profiles of the position angle $\Gamma_{phot}$ and the flattening $q_{phot}$ using the kinemetry code optimised for surface photometry.  We compared the results of the moment of inertia method with the averages of the rings between the sky level and a level at 6 times the sky rms. The standard deviation of the differences between the two estimates for PA$_{phot}$ and $\epsilon$ were $2\degr$ and 0.03, respectively. A special care should be given to the estimate of $\epsilon$, particularly when $\epsilon \sim 0$, as ellipticity values are bound ($> 0$),  which induces a positive bias at low ellipticities. We estimated this bias by constructing round models with de Vaucouleurs' profiles, using brightness, noise patterns and sky backgrounds similar to the observed galaxies. Our tests suggest that the moments of inertia method affects the estimate of $\epsilon$ by a positive bias of about 0.02.

\subsection{Uncertainties on position angle estimates}
\label{ss:errors}

The uncertainties for both PA$_{kin}$ and PA$_{phot}$ are, generally, small, as can be seen from  Fig.~\ref{f:errorPA}. In the case of photometry (upper panel) there is an expected trend of larger errors with decreasing ellipticity, since PA$_{phot}$ is not a defined quantity for a circle. For $\epsilon > 0.4$ the mean measured uncertainty is just under $2\degr$, while for $\epsilon< 0.4$ it increases to just above $9\degr$. Most of the galaxies with larger uncertainties are either barred/ringed, interacting or dusty systems. Similarly, in the case of kinematics (lower panel), there is a clear trend of increasing errors with decreasing maximum rotational velocity observed within the SAURON field-of-view. The average uncertainty on PA$_{kin}$ for systems with $k^{max}_{1}> 100$ km/s is just above $3\degr$, while for $k^{max}_{1}< 100$ km/s the mean error is $17\degr$. The existence of bars/rings or evidence for interaction does not influence the accuracy of PA$_{kin}$ determinations. Dust has some influence, but it is the disappearance of rotation that causes the large uncertainties in PA$_{kin}$.

Measurements by the moment of inertia method can be systematically biased by dust obscuration, interaction features (shells, tidal streams, accreted components), morphological features (bars, rings) and bright stars or companion galaxies.  Most of our galaxies are dust free and when present, dust is mostly centrally distributed. Bright stars can be avoided in most cases by masking, which usually also works well on companion galaxies unless the pairs are very close. On the other hand,  going out to large scales to avoid bars, increases the probability to detect shells and brighter tidal debris in other galaxies. It is possible to avoid both problems if there is no {\it a priori} set radius at which (PA$_{phot}$, $\epsilon$) are measured, but an optimal one is chosen for each object instead. This, however, has to be taken into account during the analysis of the data, and could be revised for different purposes. 

\begin{figure*}
         \includegraphics[width=\textwidth]{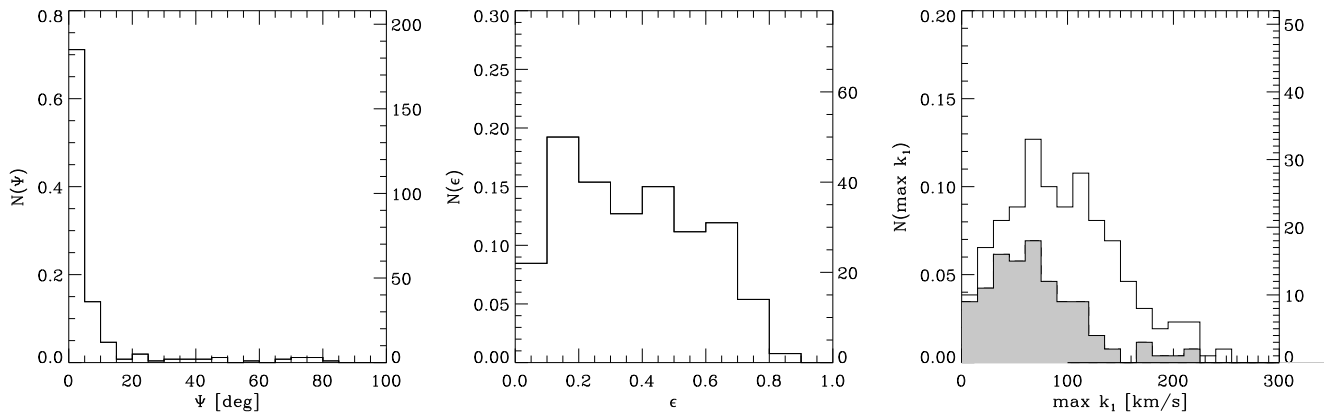}
\caption{\label{f:hist}  {From left to right:} Histograms of the kinematic misalignment angle, ellipticity and maximum rotational velocity. The left hand y axis is normalised to the total number of galaxies, while the right hand y axis gives the number of objects in each bin. In the right most histogram, the shaded region is for galaxies with $\epsilon \le 0.3$.}
\end{figure*}

In the case of PA$_{kin}$, the main sources of systematic errors lie in contamination by foreground stars, dust lanes, large bin size or bad bins. The stars or bad bins can be masked leaving enough information to determine PA$_{kin}$, while the dust affects the overall velocity extraction. Dusty galaxies are uniformly distributed over the parameter ranges plotted in Fig.~\ref{f:errorPA}, and there is no evidence the dust is affecting our measurements significantly. Large bins, however, mean a simple loss of spatial resolution and a degradation in the PA$_{kin}$ precision. We estimate that in about 5\% of galaxies, PA$_{kin}$ might be affected to some degree by the lower spatial resolution. This effect is accounted for in the quoted uncertainties.

Finally, the estimated uncertainties do not fully reflect the actual radial variation of the position angles; the extent of isophotal or kinematic twists are only partially represented with our derived uncertainties, especially in the case of photometry where the measurements are biased to the larger radii. On Fig.~\ref{f:PAcompre} we show a measure of the position angle twists for both photometric and kinematic data from the kinemetry analysis of the images and the velocity maps. They were estimated as the standard deviation within 1 R$_e$ or the SAURON FoV, $\Gamma^{\rm VAR}_{phot}$ and $\Gamma^{\rm VAR}_{kin}$ for photometric and kinematic radial variations, respectively. In case of photometry (upper panel), galaxies with larger $\Gamma^{\rm VAR}_{phot}$ are typically barred, but there are also interacting systems or galaxies with dust. Note that regular, undisturbed, galaxies with larger $\Gamma^{\rm VAR}_{phot}$ mostly have small ellipticities, which is also a consequence of the degeneracy in PA determination for more round objects. 

In the case of kinematics (lower panel) radial variations are seen almost exclusively in galaxies with NRR type of rotation, which also have lower maximal rotational velocities. These galaxies typically harbour KDCs and CRCs features ($\Gamma^{\rm VAR}_{kin}$ around $90\degr$). Note that for NRR galaxies with large  $\Gamma^{\rm VAR}_{kin}$ not all values should be taken at their face values. The kinemetry results are not robust in the regime when the rotation drops below the measurement level and the ellipse parameters are poorly constrained. 

We conclude this section by taking as the typical uncertainty on PA$_{phot}$ and PA$_{kin}$ a value of $5\degr$. This is a small over-estimate for flat and fast rotating systems, while somewhat less accurate for round and slow rotating galaxies, and we note that the typical uncertainty on PA$_{kin}$ is somewhat larger than the typical error on PA$_{phot}$. We will use it as the representative uncertainty when the two measurements are combined in the next section.

%
%

\section{Kinematic misalignment}
\label{s:kinmis}

Based on the \citet{1991ApJ...383..112F} definition we calculate the kinematic misalignment angle $\Psi$ as the difference between the measured photometric and kinematic position angles:
\begin{equation}
\sin \Psi = | \sin ($PA$_{phot} - $PA$_{kin})|.
\end{equation}

\noindent In this way, $\Psi$ is defined between two observationally related quantities and it approximates the true kinematic misalignment angle, which should be measured between the intrinsic minor axis and the intrinsic angular momentum vector.  In the above parametrisation, $\Psi$ lies between 0 and 90$\degr$ and it is not sensitive to differences of $180\degr$ between PA$_{phot}$ and PA$_{kin}$. 

In Fig.~\ref{f:hist} we show histograms of three quantities for galaxies in the ATLAS$^{\rm 3D}$ sample. The kinematic misalignment angle $\Psi$, is remarkably uniform: 71 per cent of galaxies are in the first bin with $\Psi \le 5\degr$, with another 14 per cent with $5 < \Psi \le 10\degr$, and in total 90 per cent of galaxies having $\Psi \le 15\degr$. The remaining 10 per cent of galaxies are spread over $75\degr$ with a few objects per bin. Before exploring in more details below the remarkable {\it near alignment} of early-type galaxies, we note a relatively flat distribution of ellipticities and the broad distribution of the maximum rotational velocity centred at about 90 km/s. 

The distribution of ellipticities of ATLAS$^{\rm 3D}$ galaxies is different from both distributions of ellipticities of 'ellipticals' and 'spirals' measured in the SDSS data \citep{2008MNRAS.388.1321P}. Our galaxies span the ellipticity range from 0 to just above 0.8 and in that sense are similar to the apparent shape distribution of spirals. There is, however, an excess of round objects relative to the late-types and an excess of flat objects relative to the early-types from samples analysed by \citet{2008MNRAS.388.1321P}. An in-depth analysis of the distribution of ellipticities in the ATLAS$^{\rm 3D}$ sample and its inversion regarding the intrinsic shape distribution will be a topic of another paper in this series.

The distribution of maximum rotational velocities can be described as a broad distribution around 90 km/s, and a tail of objects with high velocities. Our sample is different from the sample of \citet{1991ApJ...383..112F}, where most of the galaxies have rotational velocity less than 100 km/s, with a peak at $\sim$40 km/s. This is naturally explained by the fact that their sample had galaxies with $\epsilon<0.3$, as it can be seen if we plot the histogram of k$_1^{max}$ for only those galaxies (shaded region on the last panel). 

\begin{figure}
         \includegraphics[width=\columnwidth]{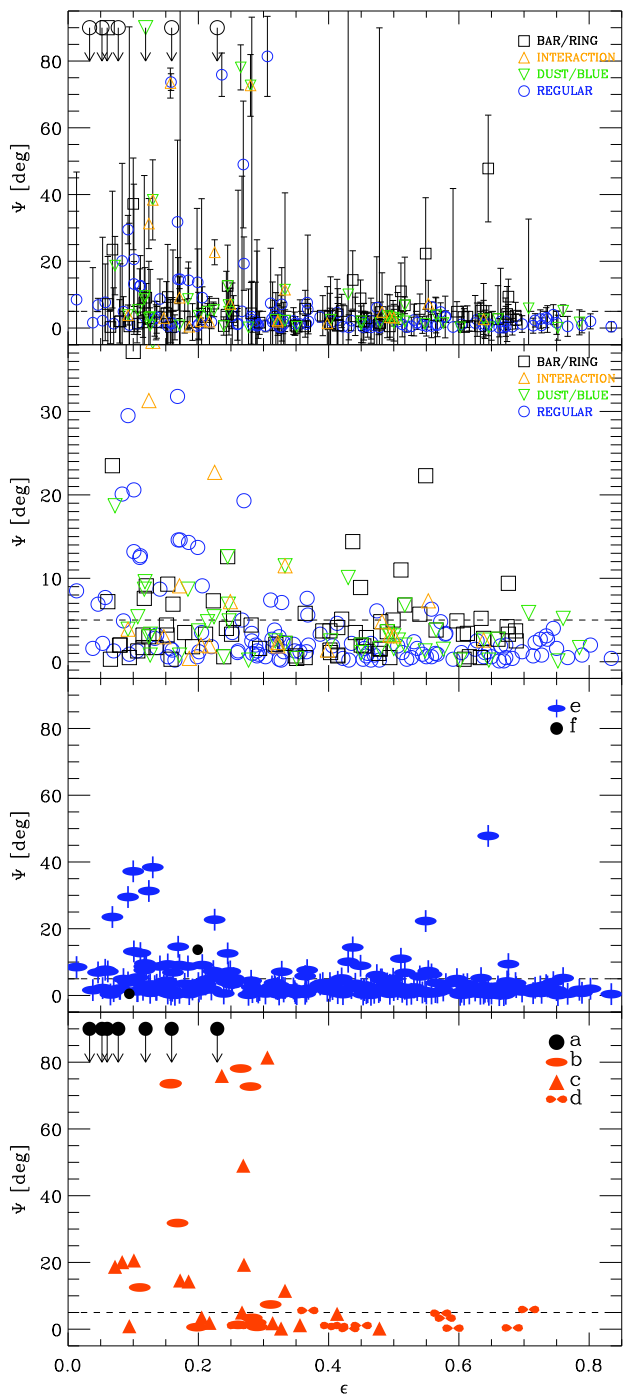}
\caption{\label{f:kinmis}  Distribution of the kinematic misalignment angle $\Psi$ as a function of ellipticity $\epsilon$. {\bf Top:} All galaxies. Different morphological features are shown with different symbols: green squares show galaxies with bars and/or rings, orange upward pointing triangles galaxies with dust or blue nuclei, red downward pointing triangles galaxies with interaction features, black circles galaxies without specific features. Large symbols without error bars show galaxies without detectable rotation (kinematic group {\it a}). The error bars are the uncertainties of the PA$_{kin}$. {\bf Middle Top:} The same plot as above, but where $\Psi$ spans only $40\degr$ and without the error bars for clarity.  {\bf Middle Bottom:} The kinematic misalignment of galaxies in kinematic groups {\it e} (regular rotators) and {\it f} (unclassified kinematics). {\bf Bottom:} The kinematic misalignment of galaxies with complex kinematic belonging to groups: {\it a} (no rotation), {\it b} (non-regular rotators without special kinematic features), {\it c} (kinematically distinct cores, including counter-rotation cores) and {\it d} ($2\sigma$ peak galaxies). On the bottom two plots the error bars are not plotted for clarity. }
\end{figure}

In the top panel of Fig.~\ref{f:kinmis} we show the kinematic misalignment angle as a function of ellipticity for all galaxies in the sample (in the second panel we show the same data, but without the error bars, and $\Psi$ in the range of $0-40\degr$). The seven larger symbols plotted as upper limits are the galaxies which do not show rotation (kinematic group {\it a}). Their uncertainties on PA$_{kin}$ are typically $\sim90\degr$, and their $\Psi$ are unconstrained, hence in this figure we plot them as "upper" limits. 

While most of the galaxies are aligned, there is a dependence of  $\Psi$ on $\epsilon$, in the sense that rounder objects are more likely to have larger $\Psi$. At the same time, however, the uncertainties increase, as shown in Section~\ref{ss:errors}.  In this section we want to scrutinise the galaxies with evidence for kinematic misalignment, and, based on the results of Fig.~\ref{f:errorPA}, we look in more details only at galaxies with $\Psi>15\degr$.

In the two top panels of Fig.~\ref{f:kinmis} we also highlight the positions of galaxies with different morphological features. Most of the galaxies with resonance phenomena have small $\Psi$.  The five most misaligned galaxies are: NGC502, NGC509, NGC2679, NGC4268 and NGC4733. NGC4733 was mentioned before (see Section~\ref{sss:links}). The other four galaxies are characterised by relatively poor kinematic data quality. NGC502 and NGC2679 have similar shapes to NGC4733, bur NGC2679 also has a prominent ring.  NGC509 and NGC4268 are interesting since they are the only galaxies flatter than 0.3 with a significant misalignment. NGC4268 has evidence for a ring, while NGC509 has a peanut shape bulge. Except in the central $\sim10\arcsec \times5\arcsec$, their velocity maps are dominated by large bins with significant changes in velocity between them, which can bias the determination of PA$_{kin}$ and might explain the unusually large $\Psi$ of these flattened objects. 

Dust or blue nuclear features are present in galaxies that are generally aligned; there are four galaxies in this class with a significant misalignment: NGC3073 (see Section~\ref{sss:links}), NGC1222, NGC3499, NGC5631 and NGC5485.  NGC1222 is an interacting galaxy with complex dust features and most likely not a settled object yet. NGC3499 has a twisted dust lane which is almost perpendicular to the observed rotation. NGC5631 has a dust disk associated with the rotation of the KDC, while NGC5485 is one of two long-axis rotators\footnote{Sometimes the long-axis rotation is also called the prolate rotation. In general, the prolate rotation is characterised by the difference between the global photometric and kinematic position angles of $\sim 90\degr$.} in our sample \citep{1988A&A...195L...5W}. It also has a dust disk of $\sim27\arcsec$ in size (just smaller than the effective radius of $28\arcsec$ and fully covering the SAURON FoV), which is oriented along the minor axis making it a polar dust-disk aligned with the stellar rotation. 

Similarly, there are five strongly misaligned galaxies with interaction features: NGC474 \citep{1999MNRAS.307..967T}, NGC680, NGC1222, NGC3499 and NGC5557. Of these, all but NGC1222 and NGC3499 are characterised by shells, while these systems are also dusty. All other galaxies with $\Psi > 15\degr$ (NGC4261, NGC4278, NGC4365 NGC4406, NGC4458, NGC5198, NGC5481, NGC5813, NGC5831) have normal morphology for early-types, but they, except NGC4278 (see below), belong to kinematic groups {\it b} and {\it c}. 

In conclusion, misaligned systems often have bars, rings, dust and interaction features and there are indications that these morphological structures influence the measurements of PA$_{phot}$. They certainly highlight a complex and, in some cases, also unsettled internal structure.  Misaligned galaxies with normal morphology have complex kinematics to which we turn our attention now.

The third from the top panel of Fig.~\ref{f:kinmis} shows the kinematic misalignment angle for galaxies belonging to the kinematic group {\it e}. These are all galaxies with simple regular rotations that can be well described by the cosine law. These galaxies are evenly spread in $\epsilon$ but are mostly found with small $\Psi$ and constitute the majority of galaxies in the first bin of the kinematic misalignment histogram (left panel on Fig.~\ref{f:hist}). There are, however, a few that are strongly misaligned (in the order of decreasing $\Psi$): NGC509, NGC3499, NGC502, NGC474, NGC4278, NGC2679, NGC680 and NGC4268. Of these only NGC4278 was not previously mentioned. Although this galaxy is classified as a RR, its kinematics show some peculiar signatures \citep{1979ApJ...229..472S,1988ApJS...68..409D,1993ApJ...407..525V,2004MNRAS.352..721E}: the mean velocity is decreasing towards the edge of the SAURON FoV and we do not cover the full effective radius. In that respect the rotation that we are seeing could also belong to a large KDC covering the FoV and it could change significantly outside the covered area (see discussion in Appendix~\ref{A:cav}). This galaxy also shows a drop in the central velocity dispersion. All these suggest it is a special case and could be classified as a KDC.

The lower panel of Fig.~\ref{f:kinmis} shows galaxies with complex kinematics and there is a significant number of strongly misaligned galaxies. We show again the galaxies from the kinematic group {\it a} (no rotation) as upper limits since their actual positions in the $\Psi - \epsilon$ is unconstrained. A Kolmogorov-Smirnov (K-S) test \citep{1992nrfa.book.....P} rejects the hypothesis that the galaxies from group {\it e} on the panel above have the same distribution of $\Psi$ as the galaxies from groups {\it b, c} and {\it d} on this panel (the probability that the distributions are the same is 0.001).

Galaxies of the kinematic group {\it b} (non-regular rotators with no kinematic features) are found both among the aligned (6) and misaligned (6) objects. Some of the most misaligned objects fall in this group, such as the long-axis rotators NGC4261 and NGC5485. A similar spread in $\Psi$ is found in galaxies of the kinematic group {\it c}, which comprises KDC and CRC systems. The only somewhat misaligned CRC system is NGC4472 ($\Psi=14\degr$), while the alignment of the KDC is rare. It  happens in some of those KDC galaxies which do not have any rotation outside the core, when the rotation of the KDC is aligned with the global shape of the galaxy.

The final group of objects on this panel is group {\it d} ($2\sigma$ peak galaxies). They are all aligned systems and except in two cases they are found only at $\epsilon> 0.4$, where there are typically no misaligned galaxies. Their velocity maps are often characterised by counter-rotating components, and in terms of kinematic misalignment they are similar to CRC galaxies (but see the discussion in Appendix~\ref{A:cav}). 

We looked for dependence of kinematic misalignment angle on both the environment and the galaxy mass, but found no strong correlations. Defining the measure of the environment as the density inside a sphere containing the ten nearest galaxies Paper VII, we found no statistical difference between the $\Psi$ for galaxies in and outside the Virgo cluster (a K-S test probability is 0.192). On the other hand, galaxies with $\Psi>15\degr$ are often found in intermediate environments with the number densities ranging between 0.01 -- 0.1 Mpc$^{-3}$. The kinematic misalignment does not depend on the mass strongly, however, splitting the sample at  $10^{11.2}$ M$_\odot$ yields a K-S test probability of 0.007, suggesting that only the most massive galaxies in our sample are more misaligned then other systems. Note that the group of most massive galaxies contains the majority of galaxies for which $\Psi$ is unconstrained, (i.e. non-rotators), which were not used in the statistical tests.

%
%
\section{Discussion}
\label{s:discuss}

The two most striking findings of this work are that (i) among nearby early-type galaxies 82 per cent show ordered, regular rotators and that (ii) 72 per cent are systems with an alignment between photometry and kinematics of less than 5 degrees, while 90 per cent are consistent with this value when the uncertainties are taken into account. There are only 10 per cent of galaxies with a large misalignment ($\Psi > 15\degr$). This finding contradicts the canonical picture of early-type galaxies and in this section we discuss our results in more detail.

\subsection{Consequence of kinematic alignment of early-type galaxies}
\label{ss:align}

The axial symmetry is the rule rather than an exception among early-type galaxies, at least within one effective radus. This is in contrast with the conventional view of early-type galaxies, in particular ellipticals. Our understanding of their structure changed from considering ellipticals simple in shape, containing little gas or dust and dynamically uncomplicated one-component systems \citep[e.g.][]{1977ARA&A..15..235G} to being dynamically complex, kinematically diverse and morphologically heterogeneous \citep[e.g.][]{1982ARA&A..20..399B, 1989ARA&A..27..235K,1991ARA&A..29..239D,1994AJ....108.1567J,1997AJ....114.1771F,2000A&AS..144...53K,2004MNRAS.352..721E, 2009ApJS..182..216K}. The first systematic observations with integral-field spectrographs and the analysis of two dimensional kinematic maps confirmed the complexity of early-type galaxies, but also showed that the traditional separation into ellipticals and lenticulars is not able to distinguish the kinematic and dynamic difference among these objects \citep{2007MNRAS.379..401E,2007MNRAS.379..418C}. Specifically, half of ellipticals in the SAURON sample were kinematically similar to lenticulars and a fraction of the other half showed signatures of triaxiality. The ATLAS$^{\rm 3D}$ sample, comprising all early-type galaxies brighter than M$_K<21.5$ and within D$<$42 Mpc, is the first sample which can address this point statistically with IFS data.

The majority of early-type galaxies are still relatively simple systems (group {\it e} with 80\% of galaxies). Their velocity maps are mostly featureless and similar to those of thin disks, although they might have multiple kinematic and morphological components such as inner disks, bars or rings. Their apparent angular momenta are typically aligned with the projected minor axis of the stellar distribution, suggesting close to axisymmetric shapes. Galaxies from group {\it e}, which show misalignments, are typically barred, have dusty features (both of which can influence the measurement of the position angles) or exhibit evidence for recent interactions (i.e. they are either not fully settled systems and/or the measurements of the position angles might be biased). 

A minor fraction of early-type galaxies show complex kinematic maps and a variety of kinemetric features (groups {\it a}, {\it b}, {\it c} and {\it d} with $\sim 20$\% of galaxies). They have multiple components with appreciably different kinematic properties (e.g KDC), some show no detectable rotation, while in others the rotation is present, but it is quantitatively different (measured by kinemetry) from the regular pattern of the majority of objects. Approximately half of the kinematically complex galaxies are significantly misaligned. As in galaxies from group {e}, there are cases of dusty or interacting galaxies with large $\Psi$, but the majority of misaligned galaxies with complex kinematics seem to be morphologically undisturbed objects and the kinematic misalignment is an evidence for their triaxial figure shapes. 

The median kinematic misalignment angle for our sample is $\sim3\degr$ which is quite different from the predictions of hierarchical structure formation models \citep[e.g.][]{2002ApJ...576...21V, 2004ApJ...616...27B,2005ApJ...627..647B,2009MNRAS.400...43C,2010MNRAS.404.1137B}, although the comparison between the cosmological simulation results and the observations can not be made directly given the differences in methods, probed regions and content of simulated and observed galaxies. The comparison with the $\Psi$ values from the remnants of mergers of equal mass disks shows that smaller misalignments are found if the mergers are dissipational \citep{2006ApJ...650..791C,2009MNRAS.397.1202J}, where the increase of gas content helps to align the angular momenta with the orientations of the minor axes of the merger remnants \citep{2010ApJ...723..818H}, but it also depends on the type of orbits of the merger and the actual Hubble type of the progenitors \citep[][Paper VII]{2011Bois}. Furthermore, the remnants of unequal mass mergers are typically aligned \citep{2006ApJ...650..791C,2009MNRAS.397.1202J, 2011Bois}, and they are likely to be significant among the formation processes for the formation of the present day population of early-type galaxies.

Within 42 Mpc there are about 9\% of misaligned early-type galaxies, or less than 3\% of the total galaxy population. This suggests that the processes that result in the large $\Psi$ measured between $\sim1$ (kinematics) and $\sim3$ (photometry) effective radii can not be very important for the formation of the majority of early-type galaxies, although they are likely important at the high mass end of the galaxy distribution.

\begin{figure*}
         \includegraphics[width=\textwidth]{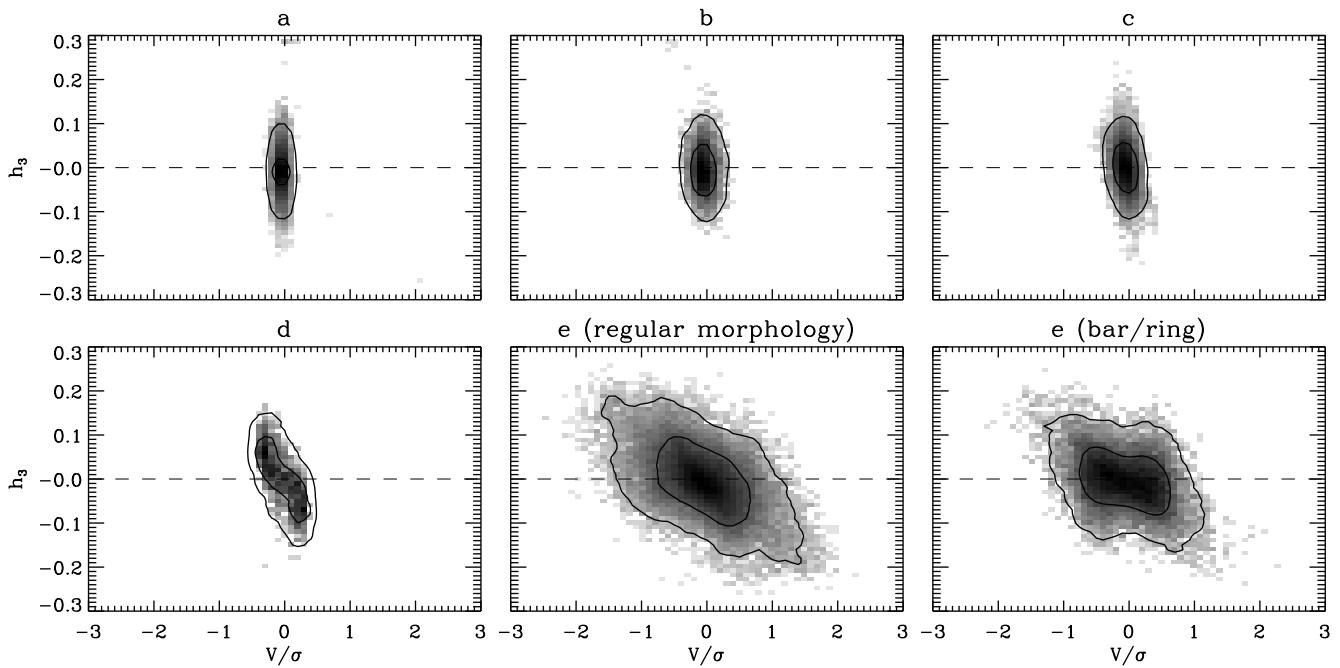}
\caption{\label{f:h3} Local $h_3 - V/\sigma$ relation for every spectrum in galaxies with $\sigma_e>120$ km/s and an error on $h_3 < 0.05$. Shown are values in bins of 0.1 in $V/\sigma$ and 0.01 in $h_3$. The colour scale is proportional to the logarithm of the intensity where the entire map sums to one. Contours enclosing 68\% and 95\% of the distributions have been smoothed using a boxcar filter and a window of 2 pixels in both dimensions. Different panels show values for galaxies separated according to their kinematics or morphology. From top to bottom, right to left:  LV galaxies (group {\it a}), NRR galaxies (group {\it b}), KDC galaxies (group {\it c}), galaxies with 2 $\sigma$ peaks (group {\it d}), RR galaxies (kinematic group {\it e}) without bars and/or rings, and RR galaxies with bars and/or rings.  }
\end{figure*}

\subsection{Disks in early-type galaxies}
\label{ss:disks}

The majority of early-type galaxies show RR type rotation, characterised by velocity maps similar to those of inclined disks ($V = V_{rot} \cos(\theta)$), having either featureless RR/NF velocity maps (66\% of the sample) or two-component RR/2M velocity maps (14\% of the sample). The vast majority of these galaxies are also kinematically aligned. Furthermore, bars and rings, which occur in disks, are found almost exclusively in galaxies with this type of rotation. 

As we show in Paper III, the division into RR and NRR types of rotation can be used to help separate the early-type galaxies into fast and slow rotators, respectively. \citet{2007MNRAS.379..418C}  and Paper III show that fast rotators, or galaxies from the kinematic group {\it e}, are consistent with being a single family of oblate objects viewed at different inclination angles. These results indicate that RR galaxies are, at least to a first approximation, made of flattened, rapidly rotating components which must be related in their origin to disks. 

Multi-wavelength observations show that gas is often present in early-type galaxies, and it is frequently settled in disks, both large H{\small I}, and small CO or ionised gas disks \citep[e.g.][]{2006MNRAS.366.1151S, 2006MNRAS.371..157M, 2008ApJ...676..317Y,2008A&A...483...57S,2010MNRAS.409..500O}. Other papers in this series will discuss these aspects in more detail, but we stress that the gas is important for the evolution of many (if not most) early-type galaxies. In addition, the stellar population content of RR galaxies often shows distinct and flattened regions of increased metallicity suggesting a link with regions of ordered rotation \citep{2006MNRAS.369..497K,2010MNRAS.408...97K}.

\begin{figure*}
         \includegraphics[width=\textwidth]{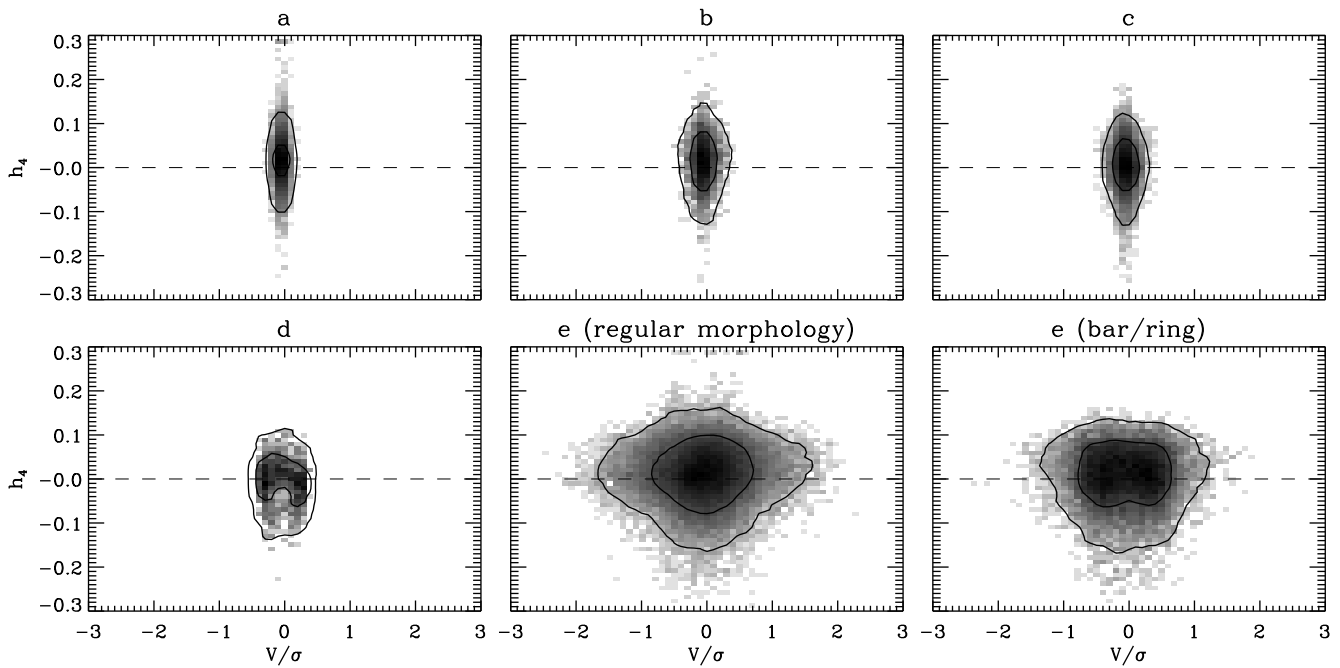}
\caption{\label{f:h4}  Local $h_4 - V/\sigma$ relation for every spectrum in galaxies with $\sigma_e>120$ km/s and an error on $h_4 < 0.05$. Different panels show values for galaxies separated according to their kinematics of morphology as in Fig~\ref{f:h3}. The colour scheme and the contours are the same as in Fig~\ref{f:h3}. }
\end{figure*}

The disk-like origin of kinematics is also visible in the higher-order moments of the LOSVD, usually parameterised by Gauss-Hermite moments, which describe the deviations from a Gaussian shape of the absorption-line profiles \citep{1993ApJ...407..525V,1993MNRAS.265..213G}. In Fig.~\ref{f:h3}, we show $h_3$ Gauss-Hermite moments separating the galaxies according to their kinematics and morphology, plotting values for each spatial bin (spectra) of those galaxies with effective velocity dispersion $\sigma_e>120$ km/s (151 galaxies). This selection is made to avoid possible biases for galaxies with $\sigma_e$ close or lower than the SAURON spectral resolution (see Paper I for details on the extraction of kinematics). The anti-correlation between $h_3$ and $V/\sigma$, which is indicative of disc kinematics \citep[e.g.][]{1994MNRAS.269..785B}, is most strongly visible in galaxies belonging to the RR kinematic class. Galaxies with the NRR type of rotation from {\it a, b} and {\it c} kinematic groups, do not show such anti-correlation, although there is a hint that among the group {\it c} galaxies (KDC and CRC) there are cases (or regions) with certain $V/\sigma - h_3$ anti-correlation. It is very interesting to see that galaxies with 2$\sigma$ peaks actually show the anti-correlation. In general, the trends are governed by the spread in $V/\sigma$ values: galaxies with the RR type of rotation have large values of $V/\sigma$, which is not the case for galaxies with the NRR type of rotation. This property is illustrated in Paper III. Note that in this respect $2\sigma$ objects are different from other galaxies with complex kinematics (groups {\it a, b} and {\it c}): the range of  $V/\sigma$ they cover is smaller than in RR galaxies, but it is bigger than for NRR galaxies. 

Bars are created from disk instabilities and it is expected that the kinematics of galaxies with bars and/or rings also show the $h_3 -$ $V/\sigma$ anti-correlation. There are, however, significant differences between RR galaxies with and without resonances: the extent of $V/\sigma$ is somewhat smaller in galaxies with bars/rings, but there is also evidence for a correlation between $h_3$ and $V/\sigma$, which can be seen in the excess of points at negative/positive $V/\sigma$ and negative/positive $h_3$ values. The existence of these correlated points is related to the correlation between the $h_3$ and $V$, typical for barred galaxies and peanut bulges \citep{2004AJ....127.3192C,2005ApJ...626..159B}

Figure~\ref{f:h4} shows $h_4$ Gauss-Hermite moments of the LOSVD for galaxies separated in the same way as in the previous figure. Again there are some differences between galaxies with the RR and NRR type of rotation. In RR galaxies for large $V/\sigma$, $h_4$ values are typically smaller and positive, but the distribution is not symmetric. This is especially noticeable for galaxies with bars/rings, while the averages of the $h_4$ distributions are, in general, slightly positive. 

The $h_3 -$ $V/\sigma$ anti-correlation is reproduced in cosmological simulation \citep{2007ApJ...658..710N}, as well as in simulations of major mergers, where the amount of gas and relative mass ratios (e.g. 1:1, 2:1, 3:1) determine shapes of the $h_3 - V/\sigma$ and $h_4 - V/\sigma$ distributions that, generally, agree well with the observations \citep{2006MNRAS.372L..78G, 2006MNRAS.372..839N}. \citet{2009ApJ...705..920H} present the latest detailed predictions for the $h_3 - V/\sigma$ and $h_4 - V/\sigma$ distributions for 1 to 1 disk mergers of varying gas fractions (from 0 to 40 per cent). The major mergers simulations reproduce some aspects of Figs.~\ref{f:h3} and~\ref{f:h4}. The $V/\sigma-h_3$ anti-correlation in gas-rich mergers (starting from 15 per cent of gas) resemble the distribution of points for galaxies of the group {\it e} without bars and/or rings. Similarly, to some extent the quantitative shape of $V/\sigma-h_4$ diagrams for large gas fractions also resembles the observations of galaxies of the group {\it e} without resonances. In both cases, however, the $V/\sigma$ range is smaller in the simulation than in the observations, with the simulations predicting overall a narrower distribution for $h_3$ and tails of positive $h_4$ values, which are not seen in the observations. It seems that the merger remnants do not rotate fast enough to reproduce the population of the RR type galaxies, but rotate too fast to reproduce galaxies with the NRR type rotation, at least within one effective radius. In contrast, the products of the consecutive dry mergers of the remnants (of the initial 1:1 mergers with 20\% and 40\% gas) better reproduce the observations, especially the fact that the $V/\sigma$ is small.

One should also keep in mind that the similarities between gas-rich merger remnants and RR type of galaxies probably come from the fact that these types of mergers produce orbital structures (e.g. short axis tubes) qualitatively similar to those allowed in nearly axisymmetric potentials of galaxies with the RR type rotations \citep{2005MNRAS.360.1185J,2010ApJ...723..818H}, but it is not clear that they create the full spectrum of observed objects among the population of early-type galaxies, suggesting that other processes should also be addressed \citep[e.g.][]{2006ApJ...636L..81N}. 

In summary, the kinemetric analysis of the velocity maps shows that the vast majority of early-type galaxies have disk-like rotation. The distribution of kinematic misalignments suggests that the great majority of early-type galaxies are nearly axisymmetric or, if they are barred, are disk systems. Their kinematic properties are only partially reproduced by equal-mass mergers. These results suggest that the disk origins of early-type galaxies remain imprinted on the entire object. Among the multiple processes that can create early-type galaxies, we need to identify a division between those that create, on the one hand, triaxial and, on the other, (close to) axisymmetric, disk dominated, remnants.

\subsection{Caveats}
\label{ss:cav}

The measured kinematic misalignment angle suggests that most early-type galaxies are nearly axisymmetric systems. This conclusion is based on the global (average) values for PA$_{kin}$ and PA$_{phot}$ not taking into account the local variations, and where scales for measuring the global values are limited by our instruments: SAURON FoV of about 1 R$_{e}$ and the SDSS imaging reaching about 3 R$_{e}$. 

If one looks at objects individually, however, a number of galaxies show local departures from axisymmetry, such as photometric and kinematic twists (e.g. $\Gamma^{\rm VAR}_{kin}$ and $\Gamma^{\rm VAR}_{phot}$  in Fig~\ref{f:PAcompre}). In addition, at least  30\% of galaxies are barred (or have bar-induced phenomena) in our sample. These objects are related to disks, but they are not axisymmetric, where the departure from axisymmetry depends on the strength of the perturbation.  Finally, the strong kinematic misalignment in about 9\% of galaxies argues for the triaxial shape of their figures. In all of these cases the internal orbital distribution is likely more complex than the one described by an exactly axially symmetric potential. 

Our kinematic measurements are confined to the central parts. It is possible that observing kinematics even further out one would start measuring larger misalignments as suggested by studies of planetary nebulae around early-type galaxies \citep{2009MNRAS.394.1249C}, although these and similar studies also find galaxies which stay (approximately) aligned \citep{2009MNRAS.394.1249C,2009MNRAS.398...91P}. It is, however, significant that we measure the kinematic and photometric position angles at different radii. To  understand the full meaning of this result it is necessary to gather kinematic observations covering a few effective radii of a larger sample of galaxies.

%
%
\section{Conclusions}
\label{s:conc}

We performed an analysis of the ground-based {\it r}-band images and the kinematic maps of 260 nearby early-type galaxies from the volume limited ATLAS$^{\rm 3D}$ sample. We used the images to determine the frequency of bars, interaction features and dust structures as well as to measure the global photometric position angle (position angle of the major axis) and the apparent ellipticity of the galaxies. 30\% of nearby early-type galaxies have bars and/or resonant rings. About 8\% of galaxies show interaction features and non fully settled figures at large radii at the surface brightness limit of the SDSS images. Barred galaxies do not show interaction features at that level of the surface brightness. We also determined local variations of these parameters using isophote fitting incorporated in the kinemetry software. Typically the global position angle and ellipticity were measured encompassing the stellar distribution within 2.5 - 3 effective radii.  

The kinematic maps are the result of SAURON observations and they consist of maps of the mean velocity, the velocity dispersion, and the $h_3$ and $h_4$ Gauss-Hermite moments. We used velocity maps to measured the global kinematic position angle (orientation of the velocity map). This angle was estimated using full maps, which typically cover one effective radius, except for the largest galaxies where they generally cover at least a half of the effective radius. 

We analysed the velocity maps applying kinemetry and used the information on the radial variation of the kinematic position angle, flattening of the maps, radial velocity profiles, and higher order harmonic terms to describe the structures on the maps and classify the galaxies according to their kinematic appearance. In doing so, we also looked for specific features on the velocity dispersion maps. This resulted in a separation of galaxies according to their rotation types:  Regular Rotators (RR) and Non-Regular Rotators (NRR). The main difference between these galaxies is that the former have velocity maps well described by the cosine law ($V=V_r \cos(\theta)$), typical for velocity maps of inclined discs. The classification was done within 1 $R_e$, or within the SAURON FoV if smaller. The ATLAS$^{\rm 3D}$ sample separates into 82\% (214) RR galaxies, 17\% (44) NRR and 2 galaxies not classified due to low quality data. This separation is used in Paper III as a basis for a separation between fast and slow rotators. The kinematic difference between RR  and NRR galaxies are also seen in the dependence of the higher order Gauss-Hermite moments ($h_3$ and $h_4$) on $V/\sigma$.

Using kinemetry we characterised various kinemetric features visible on the mean velocity maps and the velocity dispersion maps, such as: {\it No Feature} (NF), {\it Double Maxima} (2M), {\it Kinematic Twists} (KT), {\it Kinematically Distinct Cores} (KDC), {\it Counter-Rotating Cores} (CRC), {\it Low-level Velocity} (LV) and {\it Double $\sigma$} ($2\sigma$). In principle, all features could occur in galaxies with both RR and NRR type of rotation, but we find that RR galaxies are predominantly either described as NF (171) or 2M (36), while NRR galaxies are relatively equally distributed among NF (12), LV (7), KDC (11), CRC (7) and $2\sigma$ (7). Note that there are 5 exceptions to this rule: 1 RR/CRC and 4 RR/$2\sigma$ galaxies. 

In order to systematise the various kinemetric features we group galaxies in 5 kinematic groups that encapsulate the most significant features: {\it a} (NRR/LV galaxies), {\it b} (NRR/NF galaxies), {\it c} (all KDC and CRC galaxies), {\it d} (all $2\sigma$ galaxies), and {\it e} (all RR galaxies, unless they have KDC, CRC or $2\sigma$ features). The most numerous is group {\it e} (209 galaxies) and the least numerous is group {\it a} (7 galaxies). We show that the galaxies in groups {\it a}, {\it b}, {\it c}, and {\it d} are typically found in dense regions. This result is in agreement with the morphology - density relation of Paper VII.

Based on the global values for photometric and kinematic position angles we derive the distribution of the apparent kinematic misalignment angle ($\Psi$), which is directly related to the angle between the apparent angular momentum and the projection of the short axis, and hence related to the angle between the intrinsic angular momentum and intrinsic short axis in a triaxial system. A general expectation is that a triaxial object will have a non zero apparent kinematic misalignment angle. 

Exploiting our IFS data we find that the large majority of the galaxies are nearly aligned (71 \% of galaxies have $\Psi \leqslant 5\degr$, while 90\% are consistent with being aligned taking uncertainties into account). Most of the misaligned galaxies have NRR type of rotation, or have signatures of interactions at larger radii. A few are also barred. 

The small kinematic misalignment found in the great majority of early-type galaxies implies that they are axisymmetric, although individual objects show evidence for triaxial shapes, or bars. These systems have velocity maps more similar to the spiral galaxy disks than to the remnants of equal mass mergers. The latter appear to contribute to the formation of only a minor fraction of massive galaxies in the nearby Universe. Although our results are valid for the central baryon dominated regions of nearby galaxies only, we conclude that the formation processes most often result in disk-like objects that maintain the (nearly) axisymmetric shape of the progenitors. Candidate processes for forming the large fraction of early-type galaxies therefore include minor mergers, gas accretion events, secular evolution and environmental influences. Much less frequently the formation process produces an object with a triaxial figure. Most likely this involves major mergers with or without gaseous dissipation. The division of galaxies into RR and NRR types and the kinematic groups, can be used to infer the formation process experienced by a particular object.

\vspace{+1cm}
\noindent{\bf Acknowledgements}\\

\noindent MC acknowledge support from a STFC Advanced Fellowship PP/D005574/1 and a Royal Society University Research Fellowship. RLD acknowledges travel and computer grants from Christ Church, Oxford and support from the Royal Society in the form of a Wolfson Merit Award 502011.K502/jd. RLD also acknowledges the support of the ESO Visitor Programme which funded a 3 month stay in 2010. MS acknowledges support from a STFC Advanced Fellowship ST/F009186/1. NS and TD acknowledge support from an STFC studentship. RMcD is supported by the Gemini Observatory, which is operated by the Association of Universities for Research in Astronomy, Inc., on behalf of the international Gemini partnership of Argentina, Australia, Brazil, Canada, Chile, the United Kingdom, and the United States of America. TN acknowledges support from the DFG Cluster of Excellence: "Origin and Structure of the Universe". The authors acknowledge financial support from ESO. This work was supported by the rolling grants `Astrophysics at Oxford' PP/E001114/1 and ST/H002456/1 and visitors grants PPA/V/S/2002/00553, PP/E001564/1 and ST/H504862/1 from the UK Research Councils. This paper is based on observations obtained at the William Herschel Telescope, operated by the Isaac Newton Group in the Spanish Observatorio del Roque de los Muchachos of the Instituto de Astrof\'{\i}sica de Canarias. This project made use of the IDL Astronomy User's Library (http://idlastro.gsfc.nasa.gov/) \citep{1993adass...2..246L}.  We acknowledge the usage of the MPFIT routine by \citet{2009ASPC..411..251M} in \textsc{kinemetry} . This research has made use of the NASA/IPAC Extragalactic Database (NED) which is operated by the Jet Propulsion Laboratory, California Institute of Technology, under contract with the National Aeronautics and Space Administration. We acknowledge the usage of the HyperLeda database (http://leda.univ-lyon1.fr). Funding for the SDSS and SDSS-II was provided by the Alfred P. Sloan Foundation, the Participating Institutions, the National Science Foundation, the U.S. Department of Energy, the National Aeronautics and Space Administration, the Japanese Monbukagakusho, the Max Planck Society, and the Higher Education Funding Council for England. The SDSS was managed by the Astrophysical Research Consortium for the Participating Institutions. This publication makes use of data products from the Two Micron All Sky Survey, which is a joint project of the University of Massachusetts and the Infrared Processing and Analysis Center/California Institute of Technology, funded by the National Aeronautics and Space Administration and the National Science Foundation.

\bibliographystyle{mn2e}


\appendix

\section{Kinemetric analysis}
\label{A:cav}

When using the kinemetric analysis one has to be aware of the instrumental and method related sources of systematic errors. An in-depth description of the method and its application on velocity maps of early-type galaxies are presented in \citet{2006MNRAS.366..787K,2008MNRAS.390...93K}. Here, we briefly review the main sources of systematic errors. The instrumental effects come from the spatial coverage, or the size of the field-of-view (FoV), and the spatial resolution.  They particularly influence the recognition of the large and small scale kinematic structures. The SAURON pixel scale is 0.8\arcsec with a typical seeing of 1.5\arcsec (full-width-half-maximum), and the nuclear structures of comparable sizes are not likely to be detected. This, in particular, affects KDC, CRC and 2M kinemetric features. For example, observations with OASIS, an IFS with higher spatial resolution, showed that the nuclear regions of NGC4150 and NGC4621 actually contain small CRC \citep{2006MNRAS.373..906M}. 

On the other hand, for some galaxies the FoV of our observations did not cover fully one effective radius. It is possible that a full coverage (up to 1 R$_e$) of some galaxies would reveal, more generally, a different type of rotation, or, more specifically, a certain kinemetric feature. For example, NGC3607 or NGC4278 are classified as RR galaxies, but having a full 1 R$_{e}$ coverage one may characterise them as NRR/KDC galaxies. 

The effects intrinsic to kinemetric analysis are related to the assumption that a velocity map is an odd moment of the LOSVD. In other words, that there is a detectable rotation and that there are receding and approaching parts of the map. In order to constrain the parameters of the best fitting ellipse ($\Gamma_{kin}$ and {\it q}) it is necessary that the velocity map resembles to some extent the classical spider diagram. If there is no rotation, if the velocity map is noisy, in the sense that there is a large variation in velocity between adjacent bins, and if the map is described by cylindrical rotation (parallel iso-velocities),  $\Gamma_{kin}$ and/or {\it q} will not be determined robustly, either becoming fully degenerate or just poorly determined. A particular consequence of this is that disk galaxies seen face on (at an  inclination of nearly $0\degr$) could be misclassified as having the NRR type of rotation and, specially, as NRR/LV galaxies. Systems with stellar disks and significant amount of dust could be particularly susceptible to this problem. They, however, are rare in our sample. Indeed, there is evidence that only three galaxies (NGC3073, NGC4733 and NGC6703) might be misclassified in this way. 

During characterisation of kinemetric features we strictly followed the prescription given in Sections~\ref{sss:class} and~\ref{sss:features} and we did not correct afterwards for the possible misclassifications mentioned above. We estimate that the largest relative contamination is indeed in the case of LV features, simply because of their low number. If the three galaxies from above are removed from group {\it a}, there would only be 4 (1.5\%) non-rotators, making these object even more rare in the local Universe.

\section{Remarks on the differences between 2M, KDC, $2\sigma$ and CRC galaxies}
\label{A:rem}

\begin{figure}
        \includegraphics[width=\columnwidth]{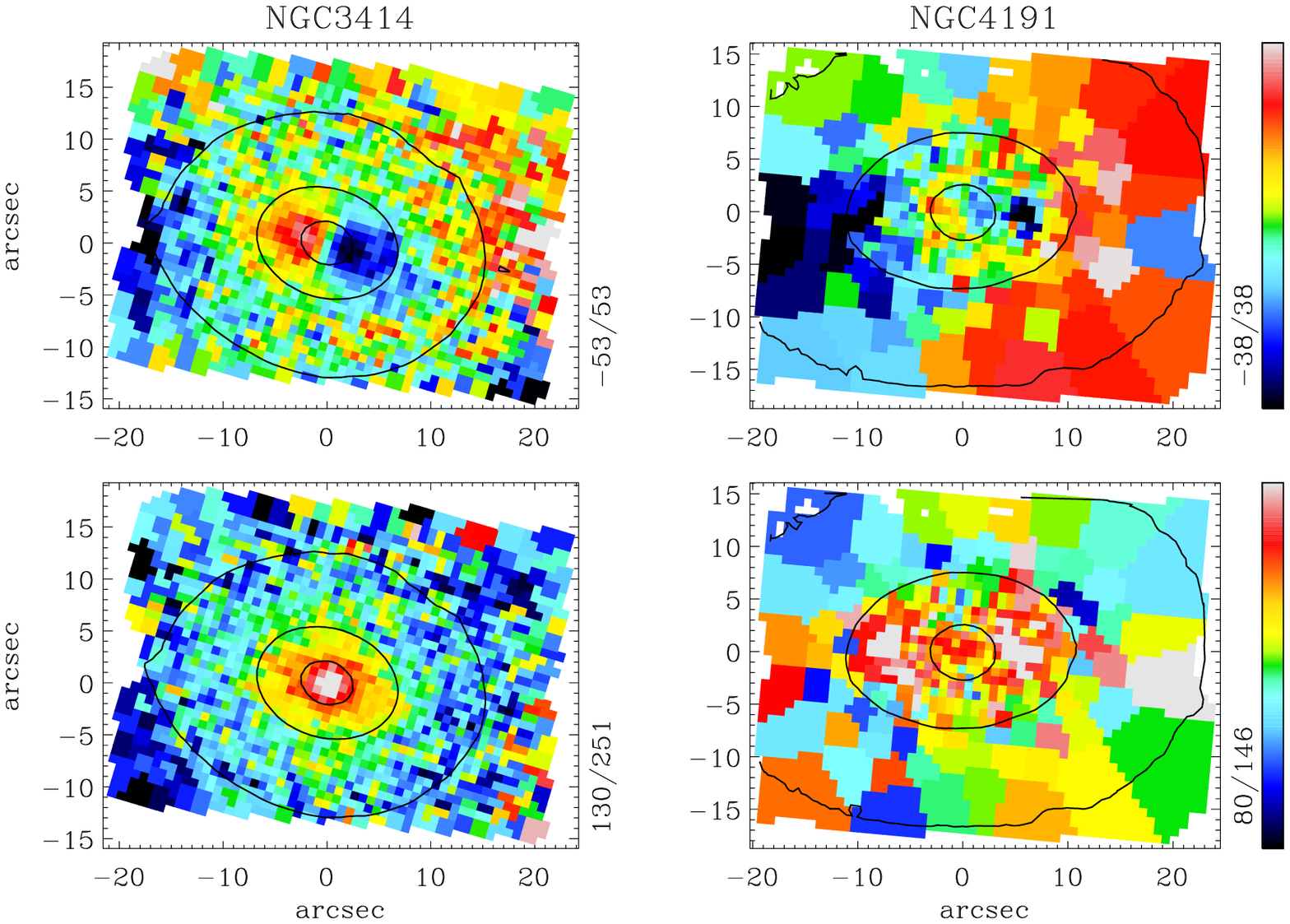}
\caption{\label{f:2s} The mean velocity {\bf (top)} and the velocity dispersion maps {\bf (bottom)} for NGC3414 {\bf (left)} and NGC4191 {\bf (right)}.  These galaxies have similar apparent shapes (0.23 and 0.27, respectively) and both have counter-rotating components on the velocity maps. Their velocity dispersion maps are very different and NGC3414 is classified as NRR/CRC while NGC4191 as NRR/$2\sigma$ galaxy.}
\end{figure}

There are two pairs of kinemetric features which deserve more attention, especially in terms of differentiation between them. They are: 2M and KDC, and $2\sigma$ and CRC. The velocity maps with 2M feature could be considered consisting of a kinematically distinct component in the central region (core) and an outer component, suggesting they are actually a subclass of KDC that happen to be aligned and show RR type rotation. They are, however, significantly different from the true KDC features. Firstly, if they would be a sub-class of KDCs than it can be expected that there should be approximately the same number of 2M and CRC galaxies (CRCs are also a subclass of KDC which is misaligned for $180\degr$ and hence a direct opposite to 2M). This is not true since there are 36 2M and 7 CRC galaxies. In addition, more than half of 2M galaxies (20) occur in galaxies with bars and/or rings phenomena, which is not the case for KDC and CRC features. This indicates that the formation scenario is different for 2M and KDC galaxies. 

Unlike all other kinemetric features, 2$\sigma$ galaxies are recognised by looking at the velocity dispersion maps. The reason is that the velocity maps of galaxies with this feature have various appearances. The most common feature on the velocity maps are counter-rotating components (e.g. NGC448), but it is possible to have multiple sign reversals (e.g. NGC4528), or ordered RR rotation (e.g. NGC4473), or even no rotation in the central region (e.g. NGC4550). The two peaks on the velocity dispersion maps which are aligned and occur on the major axis of the galaxies, are, however, always present.  The velocity dispersion maps of, for example, galaxies with the CRC features show a central increase in $\sigma$ (see Fig~\ref{f:2s} for a comparison), and, most likely CRC and $2\sigma$ galaxies have different formation scenarios. 

There is compelling evidence that the $2\sigma$ peaks are signatures of two counter-rotating disk-like structure. The most famous example of these galaxies is NGC4550 which was shown to consist of two equal in mass stellar disks with opposite angular momenta, both by studying the shape of the LOSVD \citep{1992ApJ...394L...9R,1992ApJ...400L...5R} and by constructing dynamical models \citep{2007MNRAS.379..418C}.  The latter study also showed that NGC4733, a $2\sigma$ galaxy which does not show evidence of a counter-rotation on the velocity map also consists of two components with opposite angular momenta. A similar configuration would also be the simplest explanation for the consecutive changes in velocity sign in NGC4528, as well as explain why 2$\sigma$ galaxies have both RR and NRR types of rotation. 

Most of $2\sigma$ galaxies are flattened systems seen at high viewing angles, which introduces a bias since decreasing the inclination also dilutes the signature in the velocity dispersion maps (but see \citet{2011Bois} for maps of $2\sigma$ galaxies at various inclinations), and their frequency of 4\% is likely just a lower limit. In addition, we choose to identify objects with substantial mass in the counter-rotating disks, which is reflected in the increasing separation between the two $\sigma$ peaks. There are a few galaxies which show some signatures of two peaks (e.g. NGC661, NGC4150, NGC7332), but they are not resolved well on SAURON velocity dispersion maps. 

\section{The mean velocity maps of ATLAS$^{\rm 3D}$ galaxies}
\label{A:maps}

\begin{figure*}
        \includegraphics[width=0.95\textwidth]{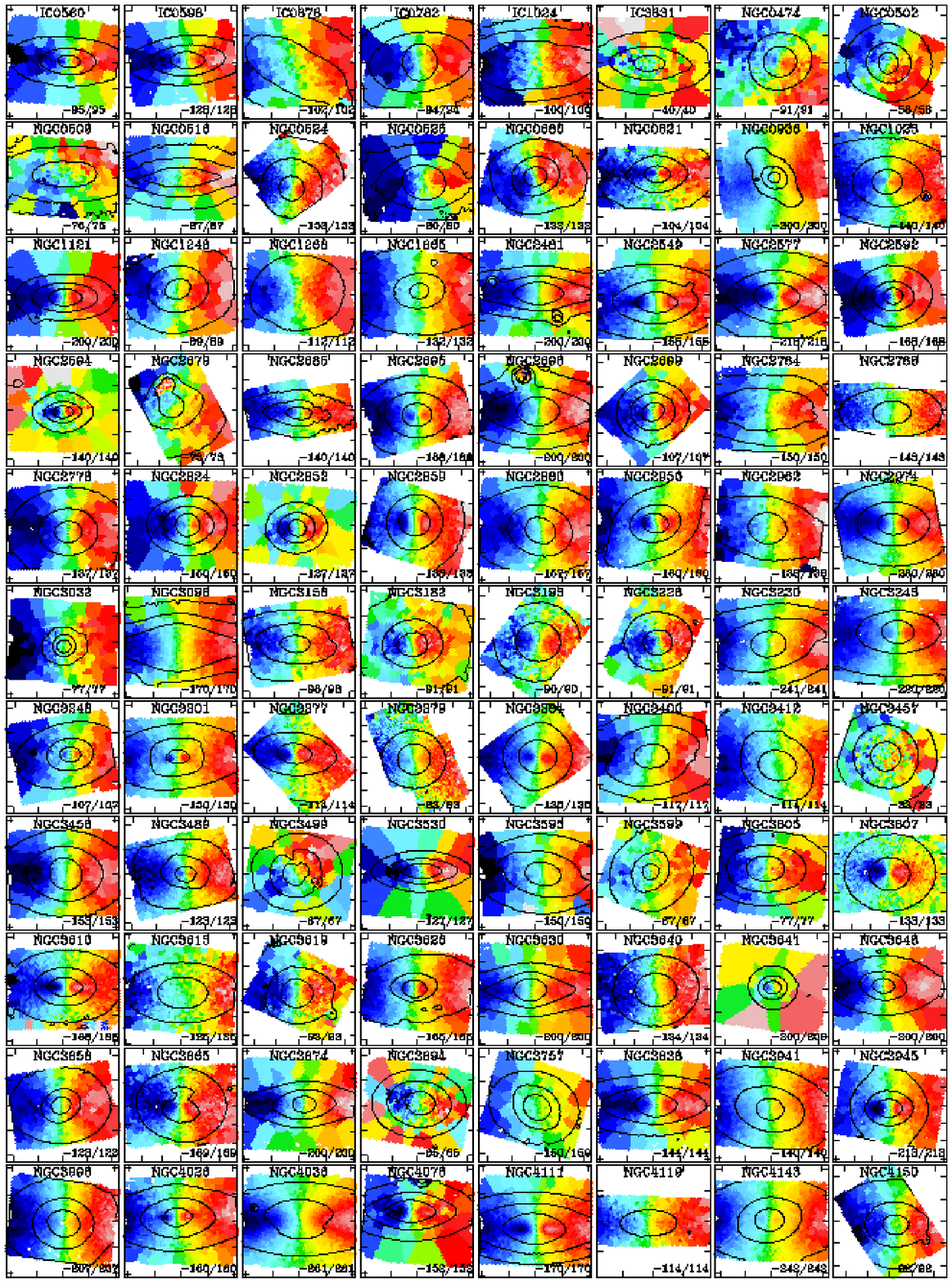}
\caption{\label{f:RR1} Velocity maps of galaxies of the kinematic group {\it e} (Regular Rotators). Contours are isphotes of the surface brightness. Maps are Voronoi binned \citep{2003MNRAS.342..345C}. All galaxies are oriented such that the global photometric axis (PA$_{phot}$) is horizontal and that the receding side is on the right. The numbers in lower right corners show the range of the plotted velocities in km/s. Ticks are separated by $10\arcsec$. Figures with maps oriented north up and east to the left are available on the project website: http://purl.com/atlas3d.}
\end{figure*}

\begin{figure*}
        \includegraphics[width=\textwidth]{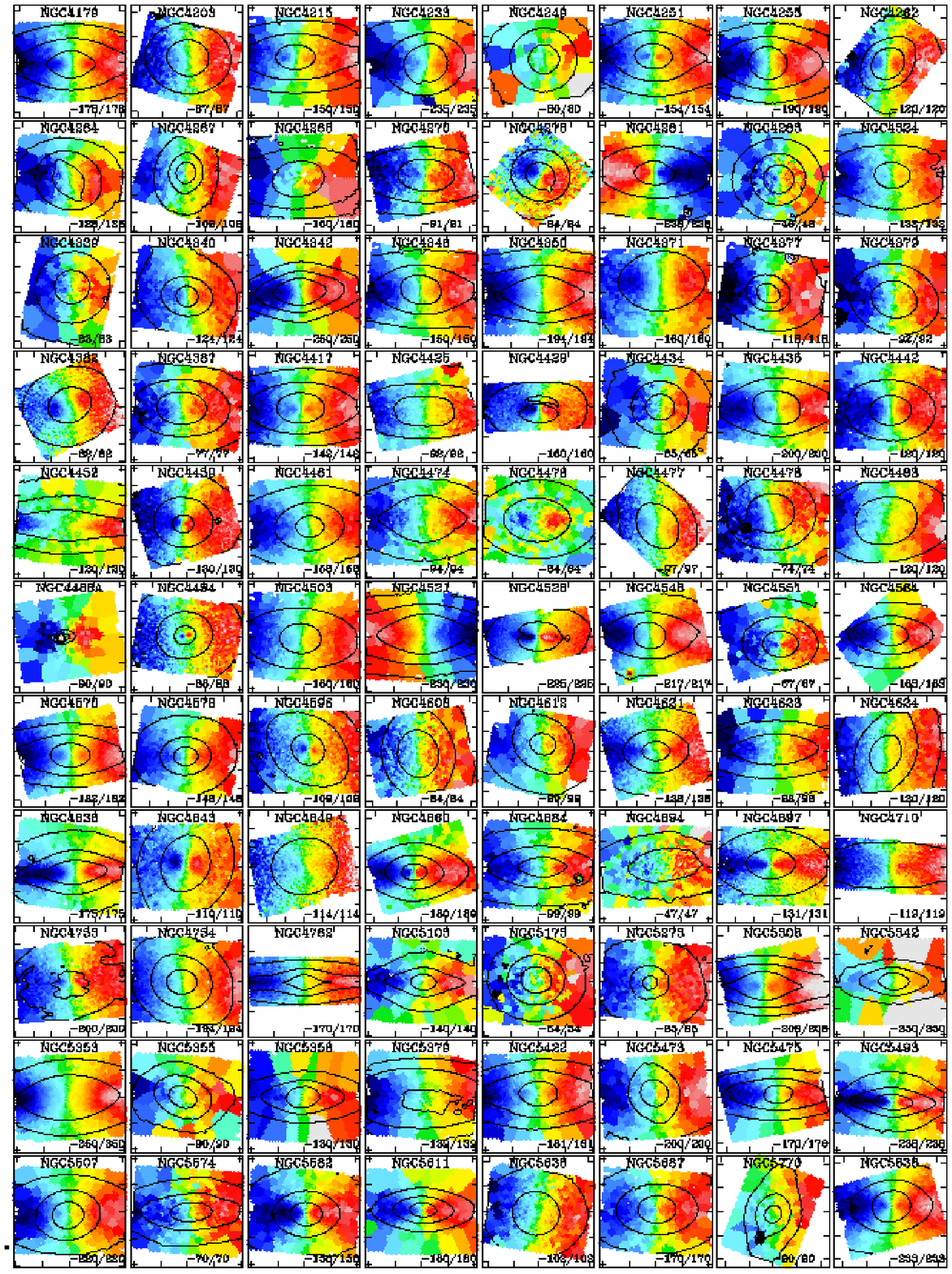}
\addtocounter{figure}{-1}
\caption{ --- continued}
\end{figure*}
\begin{figure*}
        \includegraphics[width=\textwidth]{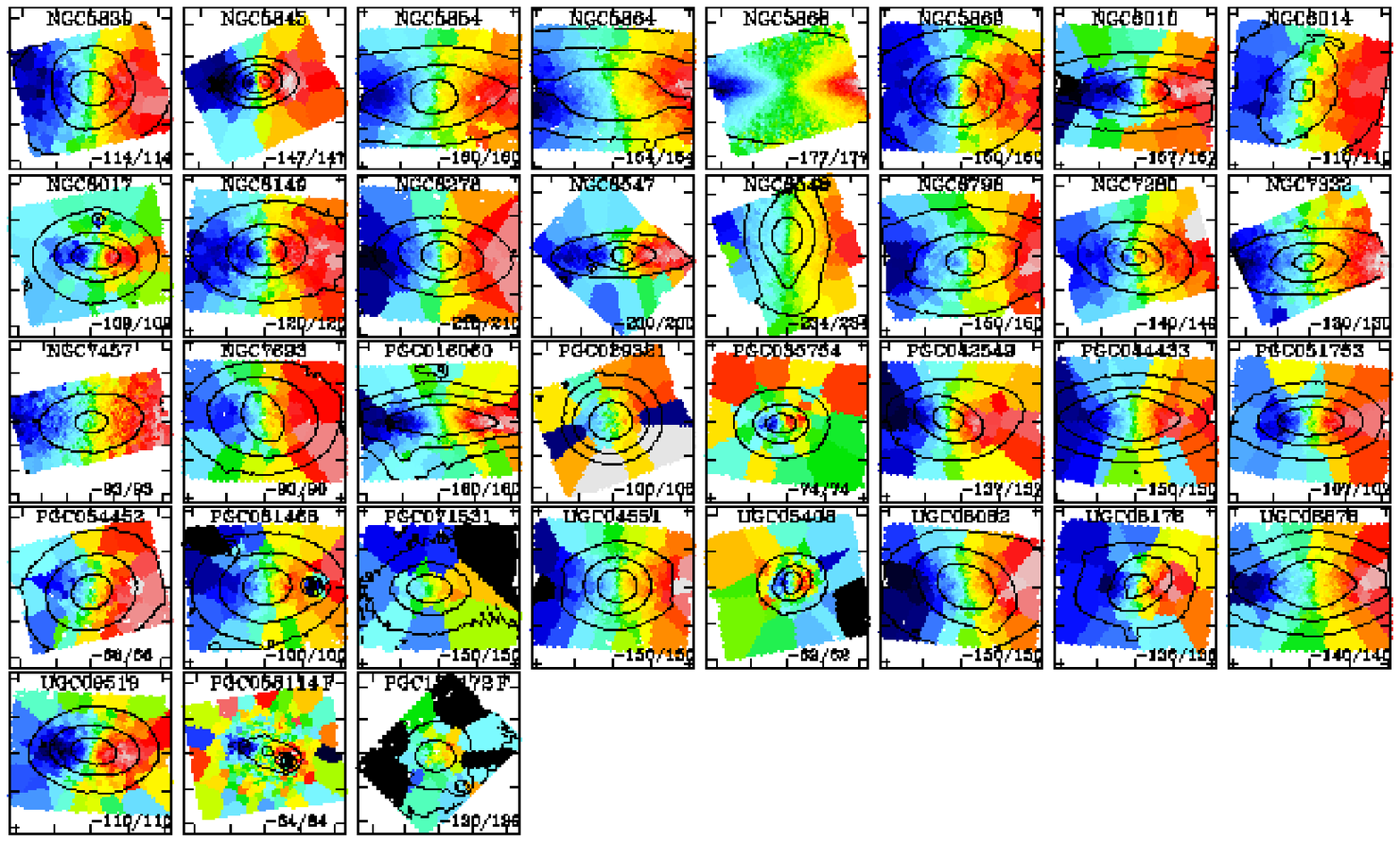}
\addtocounter{figure}{-1}
\caption{ --- continued.  Galaxies with an "F" were not classified but are plotted here for completeness. }
\end{figure*}

\begin{figure*}
        \includegraphics[width=\textwidth]{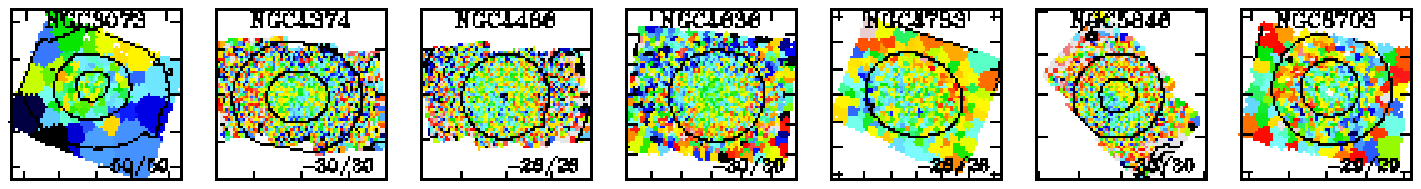}
\caption{\label{f:LV} Same as in Fig.~\ref{f:RR1}, but for galaxies of the kinematic group {\it a} (non-rotating galaxies). }
\end{figure*}
\begin{figure*}
        \includegraphics[width=\textwidth]{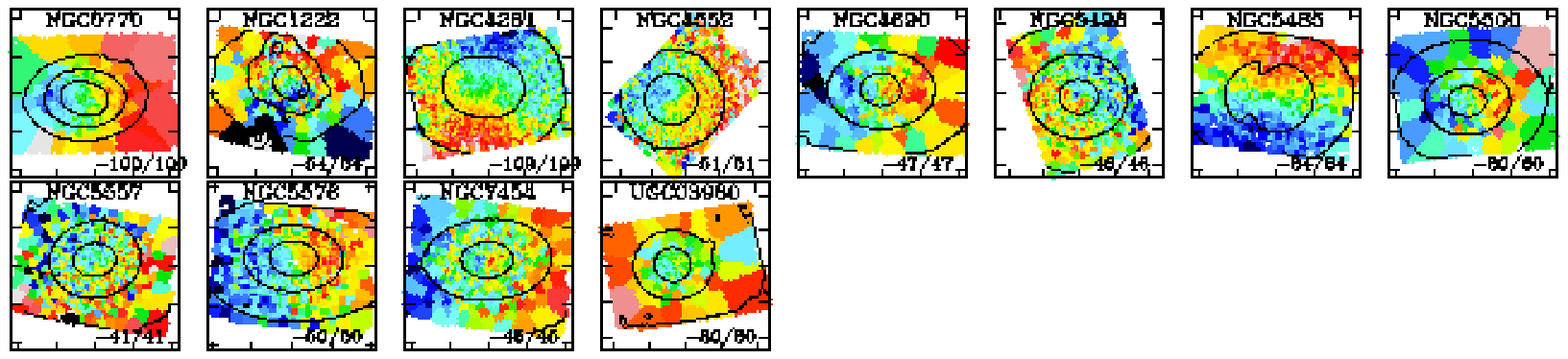}
\caption{\label{f:NRR} Same as in Fig.~\ref{f:RR1}, but for galaxies of the kinematic group {\it b} (featureless NRR galaxies).}
\end{figure*}

\begin{figure*}
        \includegraphics[width=\textwidth]{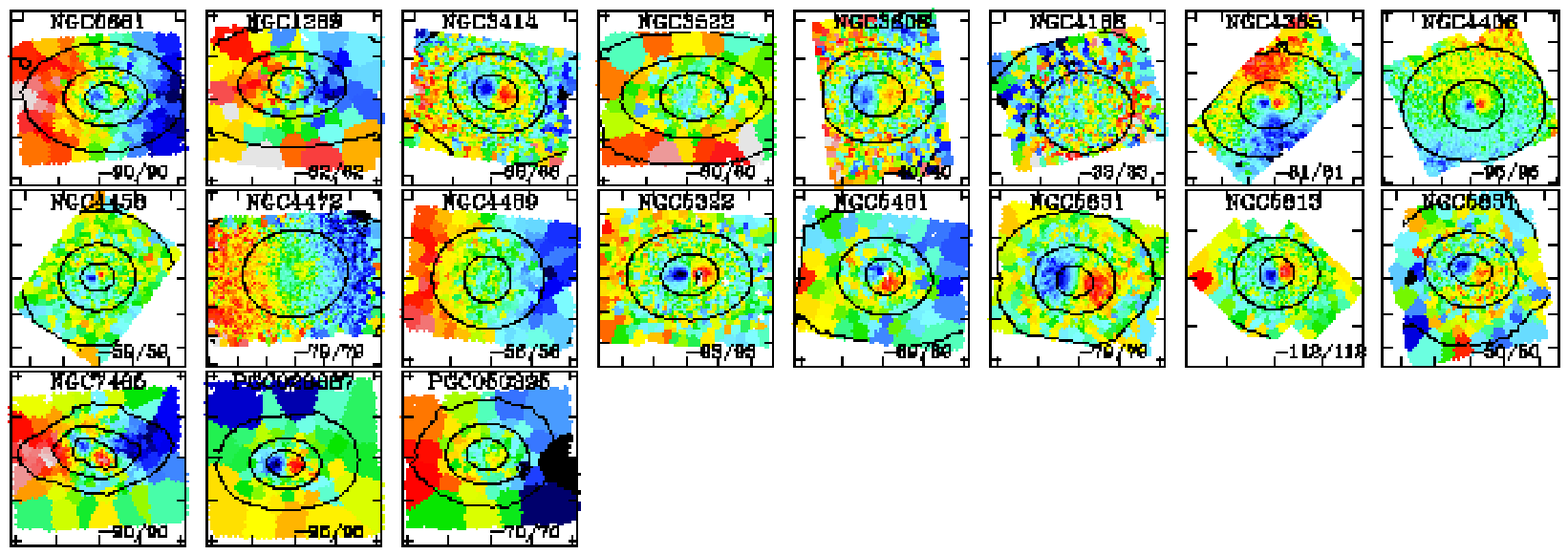}
\caption{\label{f:KDC} Same as in Fig.~\ref{f:RR1}, but for galaxies of the kinematic group {\it c} (KDC and CRC galaxies). Galaxies are oriented such that the receding side of the KDC is on the right.}
\end{figure*}

\begin{figure*}
        \includegraphics[width=\textwidth]{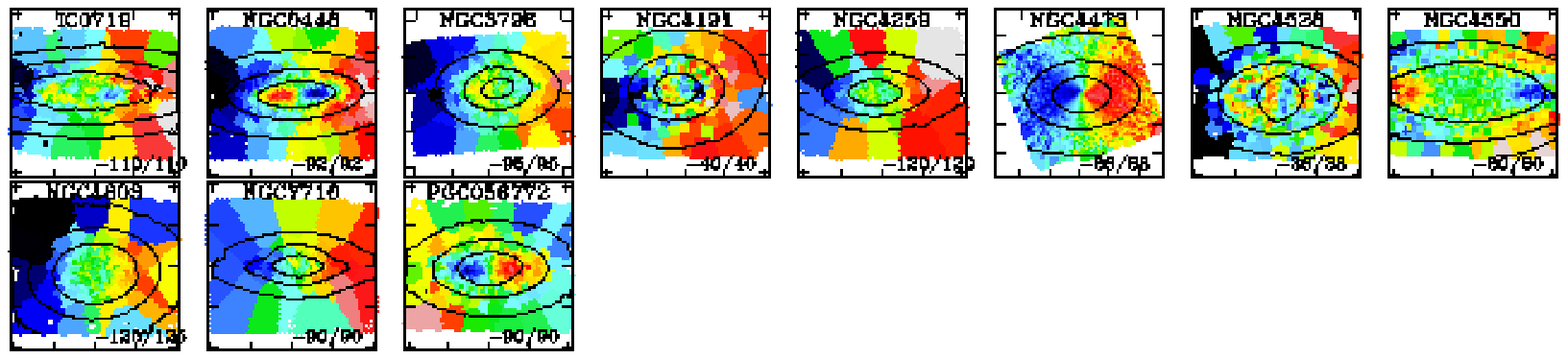}
\caption{\label{f:2sV} Same as in Fig.~\ref{f:RR1}, but for galaxies of the kinematic group {\it d} ($2\sigma$ peak galaxies).}
\end{figure*}
\begin{figure*}
        \includegraphics[width=\textwidth]{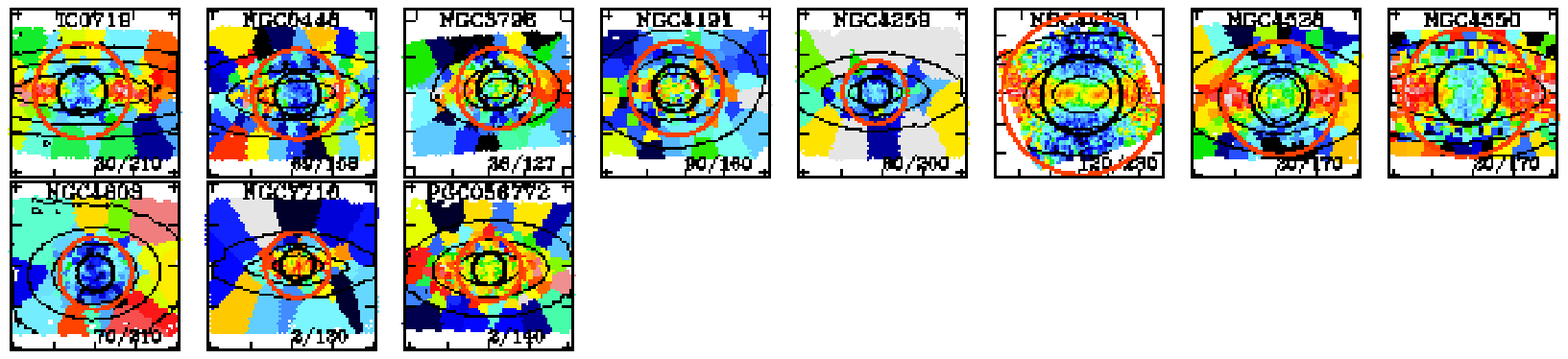}
\caption{\label{f:2sS} The velocity dispersion maps of galaxies of the kinematic group {\it d} (as as in Fig.~\ref{f:2sV}). Note two aligned peaks in the velocity dispersion which are separated by at least half of the effective radius. Over-plotted circles show one and half the effective radii. The numbers in lower right corners show the range of the plotted velocity dispersions in km/s. }
\end{figure*}

\section{Table with main properties of ATLAS$^{\rm 3D}$ galaxies used in this paper}
\label{A:master}

\clearpage
\begin{deluxetable}{l  r  r  r  c  c  c  c  c  c c}
\tablewidth{0pt}
\tabcolsep=5pt
\tablecaption{Propertied of ATLAS$^{\rm 3D}$ galaxies.}
\tablehead{
\colhead{Name} &
\colhead{PA$_{phot}$} &
\colhead{$\epsilon$               } &
\colhead{PA$_{kin}$ } &
\colhead{$\Psi$} &
\colhead{$\overline {k_5/k_1}$} &
\colhead{$k_1^{max}$} &
\colhead{Morph} &
\colhead{Dust} &
\colhead{KinStruct} &
\colhead{Group} \\
\colhead{} &
\colhead{[deg]} &
\colhead{} &
\colhead{[deg]} &
\colhead{[deg]} &
\colhead{} &
\colhead{[km/s]} &
\colhead{} &
\colhead{} &
\colhead{} &
\colhead{}\\
\colhead{(1)} &
\colhead{(2)} &
\colhead{(3)} &
\colhead{(4)} &
\colhead{(5)} &
\colhead{(6)} &
\colhead{(7)} &
\colhead{(8)} &
\colhead{(9)} &
\colhead{(10)} &
\colhead{(11)}
}
\startdata
IC0560&           18.0 $\pm$            4.1&           0.56 $\pm$           0.20&           17.0 $\pm$            8.2&            1.0&          0.036 $\pm$          0.024&           75.0&    N&    N&         RR/NF&e\\
IC0598&            4.5 $\pm$            1.4&           0.67 $\pm$           0.02&            6.5 $\pm$            4.5&            2.0&          0.026 $\pm$          0.017&          108.4&    N&    N&         RR/NF&e\\
IC0676&            5.9 $\pm$           77.2&           0.25 $\pm$           0.09&           18.5 $\pm$           12.2&           12.6&          0.034 $\pm$          0.045&           82.3&   BR&    F&         RR/NF&e\\
IC0719&           51.9 $\pm$            0.1&           0.71 $\pm$           0.01&           46.0 $\pm$           26.8&            5.9&          0.294 $\pm$          0.129&           26.9&    N&    D&        NRR/2s&d\\
IC0782&           58.1 $\pm$            2.8&           0.28 $\pm$           0.06&          242.5 $\pm$            9.2&            4.4&          0.028 $\pm$          0.022&           73.7&   BR&    N&         RR/NF&e\\
IC1024&           26.9 $\pm$            0.7&           0.64 $\pm$           0.02&           29.5 $\pm$           10.8&            2.6&          0.049 $\pm$          0.036&           71.6&    I&    F&         RR/NF&e\\
IC3631&           89.1 $\pm$            0.7&           0.43 $\pm$           0.04&           79.0 $\pm$           89.8&           10.1&          0.155 $\pm$          0.151&           15.1&    N&    B&         RR/NF&e\\
NGC0448&          114.2 $\pm$            0.2&           0.57 $\pm$           0.06&          119.0 $\pm$            2.8&            4.8&          0.050 $\pm$          0.020&           72.0&    N&    N&         RR/2s&d\\
\enddata
\tablecomments{
Column (1): The Name is the principal designation from LEDA, which is used as standard designation.\\
Column (2):  Global photometric position angle and the uncertainty in degrees, measured East of North and within 2.5 - 3 half-light radii.\\
Column (3): Global ellipticity and uncertainty, measured within 2.5 - 3 half-light radii.\\
Column (4): Global kinematic position angle and the uncertainty in degrees, measured East of North at the receding part of the velocity map.\\
Column (5): Kinematic misalignment angle in degrees. In the text, the uncertainty values for the global kinematic angle are assigned to the kinematic misalignment angle.\\
Column (6): Luminosity weighted average ratio of the harmonic terms obtained by kinemetry.  \\
Column (7): Maximal rotational velocity reached within the SAURON field-of-view.\\
Column (8): Morphological properties of galaxies: {\it B} - bar, {\it R} - ring, {\it BR} - bar and ring, {\it S - shells}, {\it I - other interaction feature}, {\it U - unknown}.\\
Column (9): Dust features: {\it D} - dusty disk, {\it F} - dusty filament, {\it B} - blue nucleus, {\it BR} - blue ring. Combinations of these are possible. \\
Column (10): Kinematic structure: See Table~\ref{t:class} for detailed explanation of all classes. \\
Column (11): Kinematic Group: {\it a} - LV galaxies, {\it b} - NRR galaxies, {\it c} - KDC and CRC galaxies, {\it d} - $2\sigma$ peak galaxies, {\it e} - all other RR galaxies, {\it f} - unclassified galaxies.   \\
Values of -1.0 in Cols. (5) and (6) are given to galaxies for which kinemetry analysis was not successful.The two galaxies with $\dagger$ did not have SDSS or INT data and we used 2MASS K-band images to determine the PA$_{phot}$ and $\epsilon$. This table is also available from our project website http://purl.com/atlas3d.
}
\label{t:master}
\end{deluxetable}


\label{lastpage}

\end{document}